\documentclass[journal]{IEEEtran}
\ifCLASSINFOpdf
\else
\fi
\usepackage{booktabs}
\usepackage{amsmath}
\usepackage{amssymb}
\usepackage{lipsum}
\usepackage{tikz}
\usepackage{graphicx,multirow}
\ifCLASSOPTIONcompsoc
\usepackage[nocompress]{cite}
\else
\usepackage{cite}
\fi
\usepackage{tikz-qtree}
\usepackage{amsmath}
\makeatletter
\newcommand{\eqnum}{\refstepcounter{equation}\textup{\tagform@{\theequation}}}
\makeatother
\usepackage{algorithm}
\usepackage{algpseudocode}
\usepackage{mathtools}
\usepackage{tabularx}
\usepackage[perpage,hang,flushmargin]{footmisc}
\begin{document}
%
% paper title
% can use linebreaks \\ within to get better formatting as desired
\title{Decentralized Edge-to-Cloud Load-balancing: Service Placement for the Internet of Things}

\author{Zeinab~Nezami, Kamran~Zamanifar\thanks{Z. Nezami, K. Zamanifar are with the Department of Computer Engineering, University of Isfahan (UI), Isfahan, Iran. e-mail: \{z.nezami, zamanifar\}@eng.ui.ac.ir.}, Karim~Djemame, and Evangelos~Pournaras\thanks{K. Djemame and E. Pournaras are with the School of Computing, University of Leeds, Leeds, UK. e-mail: \{k.djemame, e.pournaras\}@leeds.ac.uk.} \\}

\maketitle
\begin{abstract}
The Internet of Things (IoT) requires a new processing paradigm that inherits the scalability of the cloud while minimizing network latency using resources closer to the network edge. On the one hand, building up such flexibility within the edge-to-cloud continuum consisting of a distributed networked ecosystem of heterogeneous computing resources is challenging. On the other hand, IoT traffic dynamics and the rising demand for low-latency services foster the need for minimizing the response time and a balanced service placement. Load-balancing for fog computing becomes a cornerstone for cost-effective system management and operations. This paper studies two optimization objectives and formulates a decentralized load-balancing problem for IoT service placement: (global) IoT workload balance and (local) quality of service (QoS), in terms of minimizing the cost of deadline violation, service deployment, and unhosted services. The proposed solution, EPOS Fog, introduces a decentralized multi-agent system for collective learning that utilizes edge-to-cloud nodes to jointly balance the input workload across the network and minimize the costs involved in service execution. The agents locally generate possible assignments of requests to resources and then cooperatively select an assignment such that their combination maximizes edge utilization while minimizes service execution cost. Extensive experimental evaluation with realistic Google cluster workloads on various networks demonstrates the superior performance of EPOS Fog in terms of workload balance and QoS, compared to approaches such as First Fit and exclusively Cloud-based. The results confirm that EPOS Fog reduces service execution delay up to 25\% and the load-balance of network nodes up to 90\%. The findings also demonstrate how distributed computational resources on the edge can be utilized more cost-effectively by harvesting collective intelligence.
\end{abstract}
% IEEEtran.cls defaults to using nonbold math in the Abstract.
% This preserves the distinction between vectors and scalars. However, if the journal you are submitting to favors bold math in %the abstract, then you can use LaTeX's standard command \boldmath at the very start of the abstract to achieve this. Many IEEE %journals frown on math in the abstract anyway.
% Note that keywords are not normally used for peerreview papers.
\begin{IEEEkeywords}
Internet of Things, service placement, load-balancing, edge computing, cloud computing,  edge-to-cloud, fog computing, distributed optimization, collective learning, agent
\end{IEEEkeywords}

% For peer review papers, you can put extra information on the cover
% page as needed:
% \ifCLASSOPTIONpeerreview
% \begin{center} \bfseries EDICS Category: 3-BBND \end{center}
% \fi
%
% For peerreview papers, this IEEEtran command inserts a page break and
% creates the second title. It will be ignored for other modes.

\IEEEpeerreviewmaketitle
\section{Introduction}\label{sec:introduction}
\IEEEPARstart{T}{he} Internet of Things (IoT) has unprecedented impact on how data are shared and processed. IHS Markit\footnote{Available at: https://ihsmarkit.com/industry/telecommunications.html (last accessed: Jan 2021).} estimates the number of IoT-connected devices reaches 125 billion in 2030. These devices generate a large volume of data and transmit it to cloud data centers for processing, which results in the overloading of data centers and networks. However, despite several advantages of cloud computing as a shared pool of resources and services, some emerging IoT applications cannot work efficiently on the cloud. Applications, such as wind farms and smart traffic light systems, have particular characteristics (e.g., large-scale, geo-distribution) and requirements (e.g., very low and predictable latency)~\cite{buyya2019fog}.

\par Running applications in distant cloud centers results in a large and unpredictable latency. In addition, privacy and security concerns prohibit transferring sensitive data to a remote data center across the public Internet. Furthermore, due to low bandwidth, it is not efficient or even feasible to quickly transmit the high-frequency traffic generated at the network edges across the Internet. Nevertheless, compared to centralized cloud centers~\cite{bonomi2014fog}, there is a large number of distributed edge nodes with unexploited resources that can be utilized to bring low latency and reduced bandwidth to IoT networks. \emph{Edge computing} (e.g., fog computing) has been recently introduced to address the mentioned challenges by bringing the computation, storage, and networking close to the network edges where the data is being generated~\cite{Nezami2019Internet,bonomi2014fog,chiang2016fog,verma2016real,verma2015architecture,yousefpour2019all}. The introduction of recent edge computing frameworks by major cloud computing companies such as Google Cloud IoT Edge\footnote{Available at: https://cloud.google.com/iot-edge (last accessed: Feb 2021).} and Amazon AWS Greengrass\footnote{Available at: https://aws.amazon.com/greengrass (last accessed: Feb 2021).} clearly demonstrate their computation movement towards bringing cloud services closer to the edge networks. 
\par Fog computing is a system architecture that utilizes the resources along an edge-to-cloud hierarchy (i.e., fog continuum~\cite{buyya2019fog,openfog2017openfog}) to reduce traffic on the network and enhance the quality of service (QoS)~\cite{openfog2017openfog} for delay-sensitive applications. Although the federation from edge to cloud leads to new opportunities, it also raises new challenges~\cite{Nezami2019Internet,verma2015architecture}. Distribution of IoT services on available edge-to-cloud resources, which is the subject of this paper, is one of the most critical challenges concerning the federation.

\par IoT service placement is a middleware service that aims at finding one or more eligible deployments that adhere to the QoS\footnote{This paper considers QoS in terms of service execution delay and delay threshold.} expectations of the services. Placement of IoT services is a multi-constrained NP-hard problem~\cite{brogi2017}. It has a significant impact on network utilization and end-to-end delay. As a result of the dynamic nature of IoT workload~\cite{colistra2015task,brogi2017,yousefpour2019fogplan}, inefficient service placement and load-imbalance\footnote{Load-balancing refers to the distribution of workload uniformly across network resources to enhance resource utilization and network efficiency~\cite{kumar2014distributed}.} result in degradation in QoS~\cite{song2017approach,khattak2019utilization}.
A balanced distribution of workload over the network ensures the high availability of nodes to reliably forward requests to appropriate nodes~\cite{khattak2019utilization}, which reduces the probability of late response to emergencies\cite{khattak2019utilization,rahmani2015smart,gia2019exploiting}. It is also evident that the distribution of edge-to-cloud nodes, the demand for emerging distributed applications (such as smart traffic light systems~\cite{banerjee2017multi,8595391}), and the partial view that nodes have of the whole network (no centralized control)~\cite{brogi2017,colistra2015task,8875188} challenge the application of cloud central management model to this problem. Instead, this paper studies a fully decentralized management strategy, in which network nodes cooperate in allocating available resources.
\par This research introduces a decentralized load-balancing placement of IoT services in a distributed edge-to-cloud infrastructure, taking into consideration two objectives: achieving a desirable workload balance (global objective) and minimizing service execution cost (local objective). While the design of service placement algorithms for fog computing has received considerable attention in recent years~\cite{santos2019resource,alsaffar2016architecture,fan2018application}, agent-based cooperative service placement, aimed at decentralized load-balancing, is an emerging timely topic.
\par The contributions of this work to addressing the service placement problem are as follows:
(i) The introduction of a model that formalizes the IoT service placement problem in a fog computing (i.e., edge-to-cloud) infrastructure, and two objectives that aim at balancing workload over the network and minimizing the cost of service execution.
(ii) The introduction of a new methodology to locally and autonomously generate eligible deployments for IoT requests by reasoning based on local network context (i.e., system view) and the characteristics of the received requests (i.e., service view).
(iii) The applicability of I-EPOS, the Iterative Economic Planning and Optimized Selections~\cite{pournaras2018decentralized}, as a general-purpose decentralized learning algorithm, in solving the IoT service placement problem.
(iv) A comprehensive understanding of how several parameters, e.g., workload distribution method, the allowed workload redistribution level in the network (hop count) for service deployment, and network size, influence the optimization objectives.
(v) New quantitative insights about the comparison of three IoT service placement approaches in terms of QoS metrics and workload distribution.
(vi) New insights and quantitative findings on performance trade-offs. They can be used to design effective incentive mechanisms that reward a more altruistic agent behavior required for system-wide optimization.
(vii) A new open dataset\footnote{Available at https://figshare.com/articles/Agent-based\_Planning\_Portfolio/7806548 (last accessed: Jan 2021).\label{1}} for the community containing service assignment plans of agents. It can be used to compare different optimization and learning techniques as well as encourage further research on edge computing for the Internet of Things.
\par The remainder of this paper is organized as follows. Section II outlines the existing related work in the field of IoT service placement in the edge-to-cloud infrastructure. Section III formulates the IoT service placement problem. Section IV introduces our distributed service placement approach called \emph{EPOS Fog}\footnote{Available at: https://github.com/Znbne/EPOS-Fog (last accessed: April 2021)}. After this, Section V discusses evaluation results regarding the proposed approach, in comparison with \emph{Cloud} and \emph{First Fit} approaches. Finally, Section VI presents a summary, along with some open directions for future work.

\section{Related Work}\label{sec:related-work}

\par Resource provisioning and service placement are major research challenges in the field of cloud computing~\cite{cardellini2016optimal,zhan2015cloud,leitner2012cost}. Given the heterogeneity of computational resources on the edge, cloud service provisioning solutions are not easily applicable in the fog area~\cite{skarlat2017}. In this section, some of the most important recent studies on service provisioning at the edge-to-cloud computing system are discussed.
\par Souza \textit{et al.}~\cite{souza2016handling} introduce a QoS-aware service allocation for fog environment to minimize the latency experienced by services respecting capacity constraints. This objective is modeled as a multi-dimensional knapsack problem aimed at co-minimizing overall service execution delay and overloaded edge nodes (load in terms of processing capacity and energy consumption). A two-step resource management approach is presented by Fadahunsi and Maheswaran~\cite{fadahunsi2019locality}, whose goal is to minimize the response time it takes for services to get served while using as little edge nodes as possible. First, for each device, a home edge and a pool of backup edge nodes are chosen. Their objective is to find the edge nodes such that the latency between them and that device is minimum. Subsequently, IoT requested services are hosted on the allocated edge nodes guaranteeing the desired response time. Another work with the same objective as the ones above~\cite{souza2016handling,fadahunsi2019locality}, is proposed by Xia \textit{et al.}~\cite{xia2018combining}. Based on a backtrack search algorithm and accompanied heuristics, the proposed mechanism makes placement decisions that fit the objective.
\par Skarlat \textit{et al.}~\cite{skarlat2017towards} present a conceptual service placement framework for the edge-to-cloud system. Their objective is to maximize the utilization of edge nodes taking into account user constraints and is optimized using a genetic algorithm. The authors introduce the concept of fog cell: software running on IoT nodes to exploit them toward executing IoT services. In addition, an edge-to-cloud control middleware is introduced, which controls the fog cells. Also, a fog orchestration control node manages a number of fog cells or other control nodes connected to it. The latter enables IoT services to be executable without any involvement of cloud nodes.
Song \textit{et al.}~\cite{song2017approach} focus on maximizing number of services that are served by edge nodes while granting the QoS requirements such as response time. They solve the problem using an algorithm that relies on relaxation, rounding, and validation. Similar to the previous works~\cite{skarlat2017towards,song2017approach}, Tran \textit{et al.}~\cite{tran2019task} provide a service placement mechanism that maximizes the number of services assigned to edge nodes. The proposed approach leverages context information such as location, response time, and resource consumption to perform service distribution on the edge nodes. 
\par Deng \textit{et al.}~\cite{deng2016optimal} formulate workload allocation in an interplay between edge-to-cloud nodes. The trade-off between power consumption and transmission delay in the interplay is investigated and solved in approximation. Simulation and numerical results provide a useful guide for studying the cooperation between edge-to-cloud nodes. A similar approach, named Fogplan~\cite{yousefpour2019fogplan}, formulates the trade-off between monetary cost (cost of processing, deployment, and communication) and service delay in the IoT platform. Fogplan monitors the incoming IoT traffic to the edge nodes and decides when it is necessary to deploy or release a service, thereby optimizing the trade-off. Naha \textit{et al.}~\cite{naha2020deadline} propose a resource allocation method for a three-layer fog-cloud architecture that consists of the fog device, fog server, and cloud layers. In order to handle the deadline requirements of dynamic user behavior in resource provisioning, available resources are ranked based on three characteristics that include the available processing time, the available bandwidth, and the response time. Then, these resources are allocated for the received requests in a hierarchical and hybrid fashion.
In another work, Naha \textit{et al.}~\cite{naha2021multi} propose a cluster-based resource allocation algorithm for the edge-to-cloud environment to achieve the same goals as Fogplan~\cite{yousefpour2019fogplan} taking resource mobility and changes in the requirements of services after submission into consideration.
\par Kapsalis \textit{et al.} ~\cite{kapsalis2017cooperative} present a four-layer architecture that includes the device, hub, fog, and cloud layers to manage the resources in an IoT ecosystem. The hub layer acts as a mediator between the device layer and the other layers. The fog layer is responsible for service management and load-balancing that applies a score-based function to decide which host is more suitable for each service. For this purpose, the fog layer profits context information such as nodes' current utilization, battery level, and latency. Xu \textit{et al.}~\cite{xu2018dynamic} propose another load-balancing resource allocation method called \small{DRAM}\normalsize. \small{DRAM} \normalsize first allocates network resources statically and then applies service migration to achieve a balanced workload over edge nodes dynamically. Donassolo \textit{et al.}~\cite{8762052} formulate an Integer Linear Programming (\small{ILP}\normalsize) problem for IoT service provisioning, taking into consideration two objectives: minimizing deployment cost (comprising of the costs of processing, memory, and data transfer) and increasing service acceptance rate. The proposed solution uses Greedy Randomized Adaptive Search procedures~\cite{feo1995greedy}, which iteratively optimize the provisioning cost while load-balancing networked nodes. 

\par Zhang \textit{et al.}~\cite{8931659} propose a task offloading architecture in fiber-wireless enhanced vehicular edge computing networks aiming to minimize the processing delay of computation tasks.
To achieve the load-balancing of the computation resources at the edge servers, two schemes are proposed based on software-defined networking and game theory. These schemes, i.e., a nearest offloading algorithm and a predictive offloading algorithm, optimize the offloading decisions for each vehicle to complete its computation task, i.e., executing locally, offloading to Multi-access Edge Computing (MEC) server connected to roadside units, and offloading to remote cloud server. To reduce the processing time for IoT requests on local servers, Babou \textit{et al.}~\cite{babou2020hierarchical} present a hierarchical cluster-based load-balancing system. The authors propose a three-layer architecture made up of edge servers, MEC servers, and central cloud. Upon receipt of a request by a node, the system verifies whether this node has enough capacity to process the request. Otherwise, neighboring nodes, neighboring clusters, and finally cloud centers are considered to distribute the request hierarchically on the network.
\par Despite the solid contributions in the aforementioned studies on IoT service placement, the proposed approach in this paper is distinguished as highly decentralized (Novelty 1) and is designed for scalable IoT networks.
Furthermore, to the best of our knowledge~\cite{mouradian2017comprehensive}, most of the existing resource management schemes ~\cite{souza2016handling,fadahunsi2019locality,xia2018combining,skarlat2017towards,song2017approach,tran2019task, deng2016optimal,yousefpour2019fogplan,naha2021multi,naha2020deadline,kapsalis2017cooperative,8931659,babou2020hierarchical} only study one objective (e.g., load-balancing, minimizing monetary cost) in the context of IoT service provisioning. In contrast, the present research studies two opposing objectives (Novelty 2) that can be extended to account for any criteria regarding the preferences of users or service providers such as energy-saving.
Moreover, contrary to this research, the approaches~\cite{xu2018dynamic,8762052} presented for the purpose of load-balancing neglect the costs related to the deadline violation which is critical for delay-sensitive IoT services.
Table \ref{tab:rw-survey} presents an overall comparison of the related studies and the proposed work.

\begin{table}
\scriptsize
\centering
\caption{Features of the Cited Papers in the Literature in Comparison with EPOS Fog}
\label{tab:rw-survey}
\setlength{\tabcolsep}{3pt}
\begin{tabular}{p{64pt}p{40pt}p{25pt}p{25pt}p{30pt}p{23pt}}
\hline
Reference&
Heterogeneity$^{\mathrm{a}}$&
QoS&
Load-balance&
Distributed&
Multi-Objective\\
\hline%\midrule
    Souza \textit{et al.}~\cite{souza2016handling}&\checkmark&\checkmark&\checkmark&-&-\\
    Fadahunsi and Maheswaran~\cite{fadahunsi2019locality}&\checkmark&\checkmark&-&-&-\\
    Xia \textit{et al.}~\cite{xia2018combining}&\checkmark&\checkmark&-&-&-\\
    Skarlat \textit{et al.}~\cite{skarlat2017towards}&\checkmark&\checkmark&-&-&-\\
	Song \textit{et al.}~\cite{song2017approach}&\checkmark&\checkmark&\checkmark&-&-\\
    Tran \textit{et al.}~\cite{tran2019task}&\checkmark&\checkmark&-&-&-\\
    Deng \textit{et al.}~\cite{deng2016optimal}&\checkmark&\checkmark&-&-&-\\
    Yousefpour \textit{et al.}~\cite{yousefpour2019fogplan}&\checkmark&\checkmark&-&-&-\\
    Kapsalis \textit{et al.}~\cite{kapsalis2017cooperative}&-&\checkmark&\checkmark&\checkmark&-\\
    Xu \textit{et al.}~\cite{xu2018dynamic}&\checkmark&-&\checkmark&-&-\\
    Donassolo \textit{et al.}~\cite{8762052}&\checkmark&\checkmark&\checkmark&-&\checkmark\\
    Zhang \textit{et al.}~\cite{8931659}&\checkmark&\checkmark&\checkmark&-&-\\
	Naha \textit{et al.}~\cite{naha2021multi}&\checkmark&\checkmark&-&-&\checkmark\\
	Babou \textit{et al.}~\cite{babou2020hierarchical}&\checkmark&\checkmark&\checkmark&-&-\\
	Naha \textit{et al.}~\cite{naha2020deadline}&\checkmark&\checkmark&-&-&\checkmark\\
    EPOS Fog&\checkmark&\checkmark&\checkmark&\checkmark&\checkmark\\
\hline%\bottomrule
\multicolumn{6}{p{245pt}}{$^{\mathrm{a}}$The proposed solution handles heterogeneity of devices without assuming any particular type of node or network~\cite{yousefpour2019all}.}
\end{tabular}
\end{table}

\section{Problem Formulation}\label{sec:problem}
\par We define the load-balancing IoT service placement problem, as follows: given a set of IoT service requests and their requirements (e.g., processing power/CPU, memory, storage, and deadline) and a set of edge-to-cloud nodes and their capabilities (e.g., CPU, memory, and storage), find a mapping between the requests and the available nodes (i.e., service placement plan) considering two objectives: workload balancing across the edge-to-cloud nodes and minimizing the cost of service execution with minimal information about the IoT end-devices in the network. 
\par Specifically, the load-balancing requirements for IoT service placement can be demonstrated via two IoT-based application scenarios: online video broadcasting\cite{bulkan2018load} and health-monitoring systems\cite{khattak2019utilization}. Online video broadcasting is intended to provide on-demand and live video content to end-users, regardless of their location. Any unexpected peak in service requests might result in a disturbance in serving the requests and quality of experience (QoE) deterioration~\cite{bulkan2018load}. In this context, the load-balancing service placement provides the flexibility to add or remove servers as demand dictates~\cite{rajan2013survey,deng2010heat}. Moreover, avoiding overloaded and under-loaded nodes prevents peak load situations, resulting in better responses to the incoming requests of different requirements and service level agreement (\small{SLA}\normalsize).
\par An IoT-based health monitoring system, involving wearable devices, wireless body sensor networks, cloud servers, and terminals, can be used for  providing high-quality health care services~\cite{gia2019exploiting}. Wearable devices monitor vital signs, such as blood pressure, and inform responsible agencies of any abnormality or emergency.
In this system, a late notification may lead to serious consequences, such as a late response to the emergency. In addition, overburdened servers may break-down and delay urgent real-time detection~\cite{khattak2019utilization,rahmani2015smart,gia2019exploiting}. Other than a reduced service delay, the balanced distribution of workload over the network ensures high availability and sufficient capacity of nodes to reliably forward requests to appropriate nodes~\cite{khattak2019utilization}.
\par This section provides insights into the problem; first, fog computing infrastructure and IoT services are defined and then, the problem formulation is explained.

\subsection{Fog computing: infrastructure and services}\label{sec:iot-arch}
\par
\begin{figure}[t]
\centering
\includegraphics[trim=2.9cm 19.8cm 7.1cm 4.1cm, width=\columnwidth]{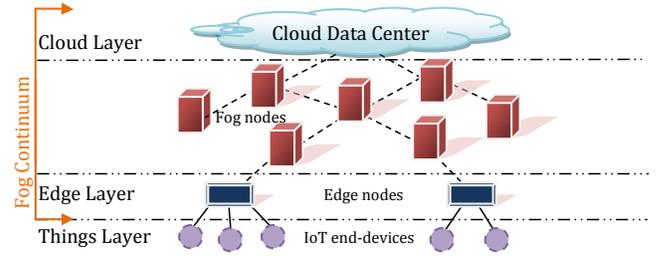}
\caption{Conceptual architecture for IoT-based fog computing networks. Fog computing occurs in the fog continuum from edge-to-cloud.}
\label{fig:iot-arch}
\end{figure}

 Fig.~\ref{fig:iot-arch} shows the principal architecture for the fog computing environment. The lowest layer of this architecture is the Things layer, where the IoT end-devices (such as mobile phones, sensors, smart wearable devices, and smart home appliances) are located. These physical devices often have low computational power and are distributed in different geographic locations~\cite{gupta2017ifogsim}. Hence, they are connected to the upper layers in order to get their services executed.
The next upper layer serves as the edge computing environment and involves edge devices such as WiFi access points and home switches.
The top layer represents the cloud computing environment, which consists of large-scale data centers and third-party cloud servers usually physically far from the Things layer.
The fog continuum, where fog computing occurs, expands from the edge network to the cloud layer (i.e., edge-to-cloud) to expand the computational capabilities of cloud computing across the fog~\cite{yousefpour2019all,buyya2019fog}. As the nodes that reside in the fog continuum cooperate as a universal system to execute services, they are referred to in this paper as fog nodes, unless stated otherwise~\cite{openfog2017openfog,iorga2018fog}.
\par An IoT application may be composed of a set of interrelated modules, i.e. a service. IoT services such as authentication and encryption are usually implemented as virtual machines (VMs) and containers that can run in different locations~\cite{yousefpour2019fogplan,skarlat2017}. Since containers share host operating system, they offer lower set-up delay compared to VMs~\cite{kaur2017container}.
IoT services can be requested from any edge-to-cloud node, and some are delay-sensitive and have tight delay thresholds (i.e., deadline), while others are delay-tolerant. Consequently, to satisfy their QoS requirements, such services may need to run close to the data sources (e.g., at the edge layer) or may be deployed further from the data sources (e.g., in the cloud)~\cite{yousefpour2019fogplan}.
Heterogeneous fog nodes, depending on their specification and capacity (e.g., processing power and storage space), could host requested services. In the present work, it is assumed that IoT services can run independently (i.e., single-service application) and are implemented in the form of containers. How the dependencies within multi-service applications affect the performance results is out of the scope of this paper and part of future work. 

\par Fig.~\ref{fig:iot-arch} shows the principal architecture for the fog computing environment. The lowest layer of this architecture is the Things layer, where the IoT end-devices (such as mobile phones, sensors, smart wearable devices, and smart home appliances) are located. These physical devices often have low computational power and are distributed in different geographic locations~\cite{gupta2017ifogsim}. Hence, they are connected to the upper layers in order to get their services executed.
The next upper layer serves as the edge computing environment and involves edge devices such as WiFi access points and home switches.
The top layer represents the cloud computing environment, which consists of large-scale data centers and third-party cloud servers usually physically far from the Things layer.
The fog continuum, where fog computing occurs, expands from the edge network to the cloud layer (i.e., edge-to-cloud) to expand the computational capabilities of cloud computing across the fog~\cite{yousefpour2019all,buyya2019fog}. As the nodes that reside in the fog continuum cooperate as a universal system to execute services, they are referred to in this paper as fog nodes, unless stated otherwise~\cite{openfog2017openfog,iorga2018fog}.
\par An IoT application may be composed of a set of interrelated modules, i.e. a service. IoT services such as authentication and encryption are usually implemented as virtual machines (VMs) and containers that can run in different locations~\cite{yousefpour2019fogplan,skarlat2017}. Since containers share host operating system, they offer lower set-up delay compared to VMs~\cite{kaur2017container}.
IoT services can be requested from any edge-to-cloud node, and some are delay-sensitive and have tight delay thresholds (i.e., deadline), while others are delay-tolerant. Consequently, to satisfy their QoS requirements, such services may need to run close to the data sources (e.g., at the edge layer) or may be deployed further from the data sources (e.g., in the cloud)~\cite{yousefpour2019fogplan}.
Heterogeneous fog nodes, depending on their specification and capacity (e.g., processing power and storage space), could host requested services. In the present work, it is assumed that IoT services can run independently (i.e., single-service application) and are implemented in the form of containers. How the dependencies within multi-service applications affect the performance results is out of the scope of this paper and part of future work. 

\subsection{Edge-cloud system model}\label{sec:system-model}
\par This section presents the notations and variables used in this paper. As shown in Fig.~\ref{fig:iot-arch}, a physical network is modeled as an undirected graph denoted by {G = (V; E)}, where V indicates the set of network nodes belonging to the different layers, and E indicates the set of edges between them. It is worth noting that {V = ({C}$\cup${F})}, where {F}  corresponds to the set of fog nodes, and C includes the cloud nodes. 
Each node $f_{j}$ is characterized by its available capacity as (i) CPU $P_{\mathtt{f},{j}}$ in MIPS, (ii) memory $R_{\mathtt{f},{j}}$ in bytes, (iii) storage $S_{\mathtt{f},{j}}$ in bytes.
Let A be a set of IoT services to be executed. Each service $a_{i}\epsilon{A}$ has to be placed on a computational resource and is defined by specific requirements in terms of deadline in millisecond and resource demands as (i) CPU $P_{\mathtt{a},{i}}$ in MIPS, (ii) memory $R_{\mathtt{a},{i}}$ in bytes, (iii) storage $S_{\mathtt{a},{i}}$ in bytes.
\par IoT service requests reach the edge layer via a local area network (LAN). The receivers, i.e. edge nodes, are responsible for deciding on where to place and execute the requests. For a requesting node of the IoT network, an edge node is defined as a switch, router, or server that acts as a gateway for the requester to connect to and is directly accessible to that node. The solution to the service placement problem is a service placement plan that contains placement decisions (i.e., binary variables), which place each service either on a fog node or on a cloud node.
The binary variables ${x}_{i,j}$, $x_{i,j'}$, and ${x}_{i,k}$ denote whether service $a_{i}$ is placed on the edge node $f_{j}$ or the fog node $f_{j'}$, i.e. other neighboring nodes, or the cloud node $c_{k}$, respectively.
$\overline{x}_{i,j}$ denotes the initial configuration of $a_{i}$ on $f_{j}$, which indicates whether $f_{j}$ currently hosts the service. These binary variables are input to the optimization problem to find future placement for requested services. The notations used in this document are listed in Table ~\ref{tab:notation}.

\par The EPOS Fog mechanism operates with no assumptions and with minimal information about IoT end-devices. There are no assumptions about these nodes, such as any specific application or protocol, virtualization technology, or hardware characteristics. The only information from IoT nodes that is required for the purpose of IoT service placement is the average propagation delay $l_{i,j}$ between the IoT end-devices and their corresponding edge nodes. Firstly, the edge nodes are usually placed near IoT nodes, which means the value of propagation delay can be omitted as stated in Section III-C1. Secondly, exact values can be calculated by the average propagation delay that approximates the round-trip delay and can be measured using a simple ping mechanism.

\begin{table}[!htb]
\setlength\tabcolsep{0pt}
\footnotesize\centering
\caption {An Overview of the Mathematical Notations}\label{tab:notation}
\smallskip
\begin{tabular*}{\columnwidth}{@{\extracolsep{\fill}}lr}
\hline
Meaning & Notation \\
\hline
    \emph{Network}\\
    Number of fog nodes&$\mid$F$\mid$\\
    Number of cloud nodes&$\mid$C$\mid$\\
    Number of network nodes&$N=$$\mid$F$\mid$+$\mid$C$\mid$\\
    Link delay between node j and $j'$&$l_{j,j'}$\\
\hline
    \emph{Cloud}\\
    Set of cloud nodes&C\\
    Cloud node k&$c_{k}$ where k $\epsilon$ \{1, \dots, $\mid$C$\mid$\}\\
    Processing capacity of $c_{k}$ (in MIPS)&$P_{\mathtt{c},{k}}$\\
    Memory capacity of $c_{k}$ (in bytes)&$R_{\mathtt{c},{k}}$\\
    Storage capacity of $c_{k}$ (in bytes)&$S_{\mathtt{c},{k}}$\\
\hline
    \emph{Fog}\\
    Set of fog nodes&F\\
    Fog node j&$f_{j}$ where j $\epsilon$ \{1, \dots, $\mid$F$\mid$\}\\
    Processing capacity of $f_{j}$ (in MIPS)&$P_{\mathtt{f},{j}}$\\
    Memory capacity of $f_{j}$&$R_{\mathtt{f},{j}}$\\
    Storage capacity of $f_{j}$&$S_{\mathtt{f},{j}}$\\
\hline
    \emph{Service}\\
    Set of services&A\\
    Service i&$a_{i}$ where i $\epsilon$ \{1, \dots, $\mid$A$\mid$\}\\
    CPU demand of service $a_{i}$ (in MIPS)&$P_{\mathtt{a},{i}}$ \\
    Memory demand of service $a_{i}$ (in bytes)&$R_{\mathtt{a},{i}}$\\
    Storage demand of service $a_{i}$ (in bytes)&$S_{\mathtt{a},{i}}$\\
    Expected response time for $a_{i}$ on $f_{j}$&$e_{i,j}$\\
    Deadline for $a_{i}$&$\tau_{i}$ \\
    Traffic arrival rate to $f_{j}$ (in MIPS)&$z_{\mathtt{f},{j}}$\\
    Processing time for service $a_{i}$ hosted on $f_{j}$&$p_{i,j}$ \\
    Waiting time for service $a_{i}$&$w_{i}$ \\
    Deadline violation of service $a_{i}$ hosted on $f_{j}$&$v_{i,j}$\\
\hline
    \emph{Binary Variables}\\
    Binary decision for $a_{i}$ on edge node $f_{j}$&$x_{i,j}$\\
    Binary decision for $a_{i}$ on fog node $f_{j'}$&$x_{i,j'}$\\
    Binary decision for $a_{i}$ on cloud node $c_{k}$&$x_{i,k}$\\
    $a_{i}$ is currently hosted on $f_{j}$&$\overline{x}_{i,j}$\\
\hline
    \emph{Plan}\\
    Number of plan generations (iteration)&t\\
    Service placement plan q&$\delta_{q}$\\
    Set of possible plans for node $f_{j}$&$\Delta_{j}$\\
    Selected plan for node j at iteration t&$\delta_{j}^{(t)}$\\
    Binary vector of possible plan $\delta$&$X_{\delta}$\\
    Utilization vector of possible plan $\delta$&$V_{\delta}$\\
    Predicted utilization variance for plan $\delta$&$\sigma_{\delta}$\\
    Realized utilization variance for plan $\delta$&${\sigma}'_{\delta}$\\
\hline
    \emph{Cost Functions}\\
    Cost of deadline violation for plan $\delta$&$O_{\mathtt{\tau},{\delta}}$\\
    Cost of service deployment for plan $\delta$&$O_{\mathtt{d},{\delta}}$\\
    Cost of unhosted services for plan $\delta$&$O_{\mathtt{u},{\delta}}$\\
    Total cost for plan $\delta$&$O_{\mathtt{c},{\delta}}$\\
    Local cost function for global plan at iteration t&$L^{t}$\\
    Global cost function for global plan at iteration t&$G^{t}$\\
    Weight controller for global and local costs&$\lambda$\\
\hline
\end{tabular*}
\end{table}

\subsection{Objectives and constraints}\label{sec:objectives}
\par This research formulates the load-balancing IoT service placement problem using two objective functions: \small{\textbf{MIN-COST}} \normalsize and \small{\textbf{MIN-VAR}}\normalsize. The \small{MIN-COST} \normalsize function aims at minimizing the cost of accomplishing requested services arising from deadlines violations, unhosted services (as the QoS requirements) and generated deployment traffic on the network. The \small{MIN-VAR} \normalsize function intends to minimize the utilization variance among network nodes as a measure of workload balance across the fog continuum. The two functions and the problem constraints are formulated below.
\subsubsection{Cost minimization}\label{sec:cost-min}
\par The service execution cost consists of three elements: number of deadline violations, number of unhosted services, and service deployment traffic. The higher the number of deadline violations and unhosted services, the lower the QoS. The more traffic imposed on the network, the lower the network performance. Equation~(\ref{eq:1}) formulates the overall cost involved in executing the service placement plan $\delta$.
\begin{equation}
O_{\mathtt{c},{\delta}}=O_{\mathtt{\tau},{\delta}} + O_{\mathtt{d},{\delta}} + O_{\mathtt{u},{\delta}}
\label{eq:1}\\
\end{equation}
where $O_{\mathtt{\tau},{\delta}}$, $O_{\mathtt{d},{\delta}}$, and $O_{\mathtt{u},{\delta}}$ specify the costs of deadline violation, services deployment, and unhosted services, respectively.
\par \textbf{Cost of deadline violation}: The response time for an IoT service is defined as the time span between the moment an end-device sends its request and the moment it receives the first response for the request. We need to check if the response time ($e_{i,j}$) for the service $a_{i}$ assigned to the fog node $f_{j}$ meets the delay threshold $\tau_{i}$ defined in \small{SLA}\normalsize. The expected response time for any service results from two metrics\cite{skarlat2017,yousefpour2019fogplan}; waiting time and processing time. Consequently, the binary variable $v_{i,j}$ indicates the violation of the service deadline as follows:
\begin{align}
v_{i,j} & = \left\{
\begin{array}{rl}
0 & \text{if } e_{i,j}<\tau_{i}\\
1 & \text{} otherwise
\end{array} \right. &e_{i,j} =& w_{i}+p_{i,j}
\end{align}
where $w_{i}$ indicates the \emph{waiting time}, which accounts for the time already passed between receiving the service request $a_{i}$ and deciding on its placement, and $p_{i,j}$ accounts for the \emph{processing time} of the request. The processing procedure in fog node $f_{j}$ for service $a_{i}$ can be viewed as an M/M/1 queuing system\cite{xiao2017qoe, Serfozo1994}. If the traffic arrival rate (in MIPS) to fog node $f_{j}$ equals to $z_{\mathtt{f},{j}}$ and the processing capacity (in MIPS) of $f_{j}$ equals to $P_{\mathtt{f},{j}}$, the computational delay (waiting time at the queue plus service time) is as follows:
\begin{align}
p_{i,j} & = \frac{1}{P_{\mathtt{f},{j}}-z_{\mathtt{f},{j}}} &z_{\mathtt{f},{j}} =& \sum_{i=1}^{|A|}P_{\mathtt{a},{i}} x_{i,j} \end{align}
The queuing system at the fog node $f_{j}$ is stable if the following constraint is met:
\begin{equation}
z_{\mathtt{f},{j}} < P_{\mathtt{f},{j}}
\label{eq:6}\\
\end{equation}

\begin{figure}[!htbp]
\centering
\includegraphics[trim=6.3cm 19.2cm 6.4cm 6.2cm, width=6.4cm]{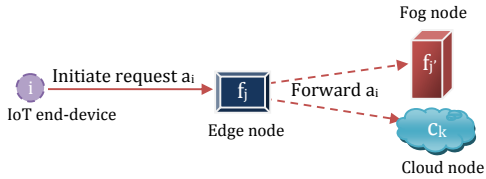}
\caption{Receiving the service request $a_{i}$ from the end-device i and forwarding it to another fog/cloud node for placement and execution.}
\label{fig:service-placement-rt}
\end{figure}
\par It is possible that the processing of a service occurs at a fog node other than an edge node (i.e., other neighboring nodes or cloud nodes). Considering Fig.~\ref{fig:service-placement-rt}, assume the end-device i is associated with the edge node $f_{j}$. The service $a_{i}$ is supposed to be executed on the neighbor fog node $f_{j'}$ or the cloud node $c_{k}$. As a consequence, the required data for processing must be transferred from i to the edge node $f_{j}$, and then, to $f_{j'}$ or $c_{k}$ for processing. Hence, we need to account for the communication delay between the end-device and the destination node. The average propagation delay between the source node i and the destination node $f_{j}$ is represented by $l(i,j)$. Note that IoT requests are input to edge nodes, which are usually located in the vicinity of end-devices, through a local area network (LAN). On the contrary, the requests are dispatched from fog nodes to cloud servers through a wide area network (WAN) that covers a large geographic area from edge to core network. Thus, the LAN communication delay could be omitted compared to the WAN~\cite{yousefpour2019fogplan,deng2016optimal}. Accordingly, the response time is formulated as follows~\cite{skarlat2017,deng2016optimal}.
\begin{equation}
\begin{aligned}
e_{i,j} = &x_{i,j}(\frac{1}{P_{\mathtt{f},{j}}-z_{\mathtt{f},{j}}}+w_{i})+x_{i,j'}(\frac{1}{P_{\mathtt{f},{j'}}-z_{\mathtt{f},{j'}}}+\\
          &2l_{j,j'}+w_{i})+x_{i,k}(\frac{1}{P_{\mathtt{c},{k}}-z_{\mathtt{c},{k}}}+2l_{j,k}+w_{i})
\end{aligned}\label{eq:7}
\end{equation}
Finally, equation (\ref{eq:8}) counts the cost of deadline violation.
\begin{equation}
O_{\mathtt{\tau},{\delta}} = \sum_{i=1}^{|A|}\sum_{j=1}^{|{C}\bigcup{F}|}v_{i,j}
\label{eq:8}\\
\end{equation}

\par \textbf{Cost of service deployment}: Deployment cost is the communication cost of service deployment, from edge to cloud nodes. When the demand for a deployed service is low, its host node may release that service to save more space. So, if a fog node accepts requests for a service not hosted locally, the service must be downloaded and deployed locally. Note that a cloud center theoretically has unlimited storage space and can host services for a long time. As a result, the communication cost for service deployment on the cloud is omitted. Equation~(\ref{eq:9}) calculates this cost~\cite{yousefpour2019fogplan}.
\begin{equation}
O_{\mathtt{d},{\delta}} = \sum_{i=1}^{|A|}\sum_{j=1}^{|F|}x_{i,j}\overline{x}_{i,j}S_{\mathtt{a},{i}}
\label{eq:9}\\
\end{equation}
where $x_{i,j}$ denotes whether service $a_{i}$ has to be placed on the node $f_{j}$, the binary variable $\overline{x}_{i,j}$ indicates if $f_{j}$ currently hosts $a_{i}$, and $S_{a,i}$ is the required amount of storage resource for deploying $a_{i}$ on  $f_{j}$.
\par \textbf{Cost of unhosted services}: If a service placement plan cannot serve all of the requests received by network nodes due to insufficient resources, this is defined as an \small{SLA} \normalsize violation. To measure this, we count the number of services that have no hosts in (\ref{eq:10}).
\begin{equation}
O_{\mathtt{u},{\delta}} = \sum_{i=1}^{|A|}(1- \sum_{j=1}^{|{C}\bigcup{F}|} x_{i,j})
\label{eq:10}\\
\end{equation}

\subsubsection{Workload balance}\label{sec:workload-balabce}
\par The second objective function consists in minimizing utilization variance among network nodes to achieve an equitable load sharing across the network. On the one hand, utilizing fog nodes can improve resource efficiency at the edge networks and help the execution of delay-sensitive services. On the other hand, load-balancing by avoiding bottlenecks (e.g., overloaded nodes) leads to a flexible network. As a result, the need for horizontal and vertical scaling up (including service migrations) due to system changes (e.g., peak times, node failures) is reduced~\cite{ningning2016fog,xu2018dynamic,banerjee2017multi}.
\par Network nodes have different capacities, and the workload allocated to them must not exceed this capacity. Thus, the workload-to-capacity proportion is applied to formulate the utilization of the nodes. Equation~(\ref{eq:11}) shows how balanced the workload distribution (in terms of processing power) is among all nodes measured with the variance:
\begin{equation}
\sigma_{\delta}=\frac{1}{|F|+|C|}\sum_{j=1}^{|{C}\bigcup{F}|}(\frac{z_{\mathtt{f},{j}}}{P_{\mathtt{f},{j}}}-\overline{\frac{z_{\mathtt{f},{j}}}{P_{\mathtt{f},{j}}}})^{2}
\label{eq:11}\\
\end{equation}
\par Note that the resource demands for a service placed on a certain node $f_{j}$ must not exceed the available resources (i.e., processing power, memory, and storage) of that node. The following three conditions ensure the capacity constraints.
\begin{equation}
\sum_{i=1}^{|A|}\chi_{\mathtt{a},{i}}x_{i,j}<\chi_{\mathtt{f},{j}}\:,\hspace{6mm} \forall f_{j}\,\epsilon\,F,\: \chi =\{P, R, S\}
\label{eq:12}
\end{equation}
Finally, the placement of services is constrained so that each service must be hosted on at most one computational resource, i.e., the fog node $f_{j}$, or the cloud node $c_{k}$. Formally,
\begin{equation}
0\leq \sum_{i=1}^{|A|}\sum_{j=1}^{|{C}\bigcup{F}|}x_{i,j}\leq |A|
\label{eq:15}
\end{equation}
Note also that the memory and processing costs in the fog nodes are assumed to be the same as the cloud. Hence, we do not account for these costs in the objective functions.

\subsubsection{Final Optimization Formulation}\label{final-problem}
\par The common way to solve the multi-objective optimization problem is modeling it as a single objective by getting a sum of the two objectives and multiplying with weight coefficients as follows:
\begin{equation}
\min (O_{\mathtt{c},{\delta}}+\sigma_{\delta}) = \\
\min (O_{\mathtt{\tau},{\delta}} + O_{\mathtt{d},{\delta}} + O_{\mathtt{u},{\delta}}+\sigma_{\delta})
\label{eq:16}\\
\end{equation}
\begin{center}
Subject to (\ref{eq:6}),~(\ref{eq:12}) - (\ref{eq:15})\\
\end{center}
The objective function of (\ref{eq:16}) is the combination of four cost functions with the same constraints which places IoT services while minimizing their execution cost and ensuring a satisfactory load-balance among network nodes. For a comprehensive study of the problem, all of the four cost functions are considered; however, some of them can be omitted in certain scenarios if needed. Note that, in some cases, minimizing a particular cost may be of higher priority in this summation. Therefore, we propose the use of a weighted scheme (i.e., $\lambda$) for modeling various preference scenarios. The next section introduces a mechanism that controls the trade-off between the two objectives.

\section{Cooperative Service Placement for IoT}\label{sec:cooperative-service}
\par This paper introduces \small{\textbf{EPOS Fog}}\normalsize, an agent-based load-balancing mechanism for IoT service placement, as the means to meet a local (individual) and a global (system-wide) objective: (i) \small{\textbf{MIN-COST}} \normalsize and (ii) \small{\textbf{MIN-VAR}}\normalsize. The former aims at reducing the cost of service execution formulated in Subsection~\ref{sec:cost-min}. Each node minimizes the cost by selecting among multiple locally generated plans for service placement that determine how service requests are assigned to the neighboring nodes. The latter minimizes the utilization variance among network nodes formulated in Subsection~\ref{sec:workload-balabce}, as a measure of load uniformity. For this purpose, nodes collaborate and exchange information with other nearby nodes to choose service placement plans that achieve load-balancing.
\par EPOS Fog is a decentralized multi-agent system that solves the placement problem. Each fog/cloud node is equipped with a software agent that autonomously generates a predefined number of \textbf{possible service placement plans} determining which service is deployed on which host in the neighborhood of the agent. Possible plans represent service placement flexibility, and each comes with a cost according to (\ref{eq:1}).
%Each agent ranks its possible plans, from low to high, according to the cost. For example, 
A plan including distant hosts (proximity in terms of hop count from source node) costs higher than a plan with closer hosts. This is because the execution of requests in further hosts imposes more traffic on the network and may result in more violations of deadlines.
\par Agents are structured in a self-organized tree topology over which they perform collective decision-making. They make coordinated selections of their possible plans considering the objectives. The process of generating and selecting placement plans repeats, agents, self-adapt their choices, and collectively learn how to optimize the objectives. Finally, the collective outcome of the choices, i.e., the \textbf{global service placement plan}, is the aggregation of the selected plans for each agent.

\subsection{Proposed solution}\label{sec:proposed-solution}
\par In the above overview, an overall understanding of the proposed load-balancing strategy has been presented. Subsequently, this subsection discusses the strategy in detail in the view of the two aforementioned objectives.
\par
\begin{figure}[!htbp]
\centering
\includegraphics[clip, trim=5.8cm 15.7cm 3.8cm 5.7cm, width=\columnwidth]{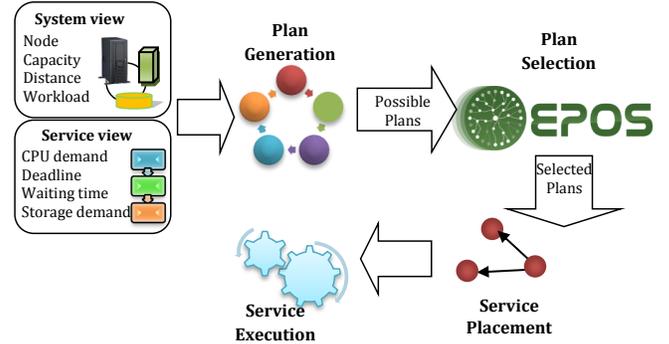}
\caption{Global system view of EPOS Fog.}
\label{fig:epos-fog}
\end{figure}

 IoT devices generate service requests and submit them to the fog nodes for placement decisions and execution. It is assumed that the receiver nodes/agents know the requirements of the received requests and the capabilities of their neighboring nodes. All receiver agents take part in a two-step procedure consists of (i) generation of local plans and (ii) plan selection. In the first step, each agent generates a set of possible plans, and in the second step, the agent selects one of them. Finally, according to the selected plan, the agents forward the received requests to the selected hosts for execution. This procedure is repeated for all new requests that enter the network. Fig.~\ref{fig:epos-fog} shows the global view of the proposed service placement mechanism, which is elaborated below.
\begin{figure}[!t]
\centering
\includegraphics[clip, trim=3.2cm 19.55cm 5.7cm 2.95cm, width=\columnwidth]{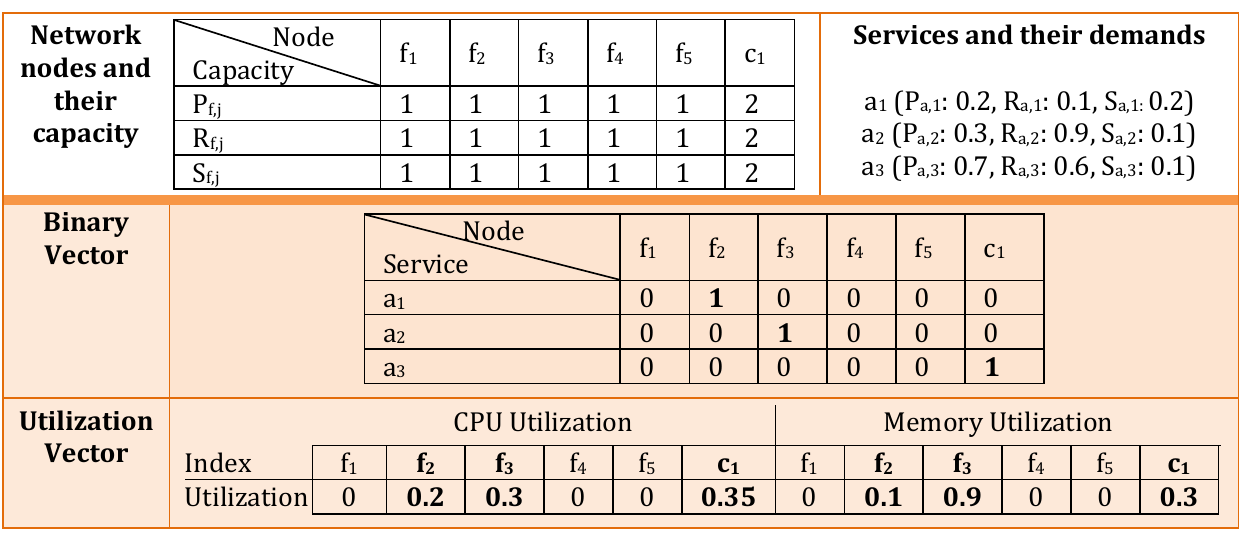}
\caption{Structure of a service placement plan for a network with five fog nodes and one cloud node.}
\label{fig:plan}
\end{figure}

\subsubsection{Generation of local plans}\label{sec:generation-plan}
\par This section illustrates how agents can locally and autonomously generate service placement plans for requested IoT services. Agents prefer to minimize their local cost, which concerns deployment traffic, service deadline violations, and unhosted services. The motivation here is that if the nodes closer to data sources (i.e., edge nodes) can be selected as hosts, deadline violations and imposed traffic on the network are minimized, resulting in higher QoS.
\par Each agent, upon receiving IoT requests, locally generates a certain number of assigning/mapping ``requests to resources'' called possible service placement plan, concerning the local cost as equation (\ref{eq:1}).
As shown in Fig.~\ref{fig:epos-fog}, each agent, for the plan generation, reasons locally based on its view of the system and requested services. For each agent, the system view represents a profile of its neighboring nodes and their features (such as capacity and distance), and the service view shows a profile of its received requests and their specifications (such as storage demand and deadline). Possible plans are the agents' options, that encode selected hosts in the form of a binary vector, and resource utilization in the form of a vector with real values.
\par The structure of a typical plan is shown in Fig.~\ref{fig:plan} that consists of a cost value and two vectors as follows. (i) \textbf{Binary vector} ($X_{\delta}$): An n-dimensional vector (n refers to the number of nodes in the network) of 0-1 values that maps the requested services to the available nodes in the network. (ii) \textbf{Utilization vector} ($V_{\delta}$): A 2n-dimensional real-valued vector that represents the resource utilization as the ratio of the assigned load to the capacity for each node. In this manner, we account for the heterogeneity in the capabilities of these nodes. Memory and CPU are considered as two metrics for the load; One half of the vector is appointed to CPU and the other half to memory. The vector can be extended to account for other metrics such as storage and battery power. However, for simplicity and to keep the size of the vector minimal, it is omitted as future work.
%Moreover, a higher size of plans increases the communication, computational and storage cost in the agents.
\par Algorithm~\ref{alg:plan-generation} illustrates the plan generation procedure. Upon receipt of requests, agents run this procedure every $\mu$ second. As a matter of design, this algorithm is a heuristically greedy algorithm that preferentially responds to the service requests which have spent a high waiting time for deployment, with respect to their deadline. Considering the distance between network nodes, we present another heuristic. Intuitively, the larger the distance of computing and communication resources, the larger the value of the local cost function. Accordingly, we design another simple rule to direct the resource allocation, i.e., the agents make a greedy decision to select closer fog resources first.
\par To generate one possible plan, each agent arranges its received requests, from low to high, in terms of the difference between the service deadline and its waiting time $\tau_{i}- w_{i}$ (line 5). Subsequently, the agent randomly chooses the required number of available neighboring nodes as candidate hosts (line 6). It then arranges these hosts ascending, according to their distance (in terms of hop count) from itself (line 7). After that, the agent assigns one to one the sorted requests to the sorted hosts while satisfying the placement constraints and taking the heterogeneity of the hosts into consideration (lines 10-16). It is assumed that up to 95\% of the capacity of each node is allocated to the requested services, and the rest is reserved for maintenance. Finally, lines 17-21 determine the service assignment to the closest cloud node if there is not enough capacity at the candidate host (line 10).
Meanwhile, utilization and binary vectors are updated accordingly. The first part of the utilization vector, i.e., CPU criterion, by lines 13 and 19, the second part of the utilization vector, i.e., memory criterion, by lines 14 and 20, and the binary vector in lines 15 and 21 are updated.
For each node, the resource utilization is measured as the ratio of the assigned load to the capacity. Note that the workload accounts for the already assigned workload (which is indicated with a bar mark in Algorithm~\ref{alg:plan-generation}) plus the new assigned workload to reach a better balance over the network. After generating a certain number of plans (which is controlled by loop for in line 3), the agent calculates the local cost for them and orders accordingly.

\begin{algorithm}{}
\caption{Local Plans Generation}\label{alg:plan-generation}
%\centering
\scriptsize{
\begin{algorithmic}[5]
\State Input:\{a: set of requested services, n: set of network nodes\};
\State Output:\{$\Delta$: set of possible plans\};
  \For{(q = 1 to $|\Delta|$)}
         \State Initialize $\delta_{q}$; $O_{\mathtt{c},{\delta}} \gets 0$;
         \State Sort a in the order of ($\tau_{i}- w_{i}$) from low to high;
         \State $h \gets$ select $|a|$ neighboring nodes from $n$;
         \State Sort h in terms of proximity from low to high;
         \State $i,j \gets 0$;
            \While {(a is not empty)}
              \State Select $a_{i}$ from a and $f_{j}$ from h;
              \If {($f_{j}$ satisfies the constraints according to inequations (\ref{eq:6}),~(\ref{eq:12}), (\ref{eq:15})}
                  \State Update $\delta_{q}$:
                  \State $V_{\delta}[j] \gets (P_{\mathtt{a},{i}}+\bar{P}_{\mathtt{f},{j}})/P_{\mathtt{f},{j}}$;
                  \State $V_{\delta}[j+n] \gets (R_{\mathtt{a},{i}}+\bar{R}_{\mathtt{f},{j}})/R_{\mathtt{f},{j}}$;
                  \State $X_{\delta}[i,j] \gets 1$;
                  \State Update the capacity for $f_{j}$;
                  \ElsIf{(the cloud node ($c_{k}$) has enough capacity)}
                  \State Update $\delta_{q}$:
                  \State $V_{\delta}[k] \gets (P_{\mathtt{a},{i}}+\bar{P}_{\mathtt{c},{k}})/P_{\mathtt{c},{k}}$;
                  \State $V_{\delta}[k+n] \gets (R_{\mathtt{a},{i}}+\bar{R}_{\mathtt{c},{k}})/R_{\mathtt{c},{k}}$;
                  \State $X_{\delta}[i,k] \gets 1$;
              \EndIf
              \State Calculate $O_{\mathtt{u},{\delta}}$ according to equation (\ref{eq:10});
              \State Remove $a_{i}$ from a and $f_{j}$ from h;
              \State i++, j++;
            \EndWhile
    \State Calculate $O_{\mathtt{d},{\delta}}$, $O_{\mathtt{\tau},{\delta}}$ according to equations (\ref{eq:8}), (\ref{eq:9});
    \State Calculate $O_{\mathtt{c},{\delta}}$ according to equation (\ref{eq:1});
    \EndFor
    \State Sort $\Delta$ in the order of $O_{\mathtt{c},{\delta}}$ from low to high;
    \State Return $\Delta$
  \end{algorithmic}
}
\end{algorithm}

The possible plans are released as open dataset\textsuperscript{\ref{1}} for the broader community to encourage further research on distributed optimization and learning for edge-to-cloud computing.
%Traffic dynamics inherent to IoT applications motivate a balanced service placement throughout the network.
\subsubsection{Plan selection}\label{sec:plan-selection}
\par IoT applications with mobility dynamics generate a varied level of network traffic~\cite{colistra2015task,brogi2017}. This motivates a balanced service placement throughout the network. As a result, more flexible assignments can be applied under various future scenarios (e.g., node failures and peak demand periods)~\cite{liu2013data,zhang2017resilient}, leading to a higher QoS and a more robust network. In this perspective, the global objective for optimizing the placement of IoT services aims at minimizing utilization variance among the network nodes, as a measure of load-balancing and peak-shaving. The utilization variance function, as shown in equation (\ref{eq:11}), is a quadratic cost function~\cite{rockafellar2000optimization} that requires coordination among agents' selections. When the autonomous agents locally generate multiple (alternative) placement plans, the placement coordination problem turns out to be a multiple-choice combinatorial optimization problem, which is NP-hard~\cite{pournaras2019socio}.
\par EPOS Fog employs the I-EPOS system\footnote{Available at: http://epos-net.org, https://github.com/epournaras/epos (last accessed: Jan 2021)}~\cite{pournaras2018decentralized,pournaras2017self}, as a fully decentralized and privacy-preserving learning mechanism for coordinating the planning of IoT requests. I-EPOS has been studied earlier in load-balancing of bike-sharing stations~\cite{pournaras2018decentralized} and demand-response of residential energy consumption\footnote{Further elaboration on the I-EPOS algorithm is out of the scope of this paper and is available on earlier work~\cite{pournaras2018decentralized}.}~\cite{pournaras2017self,pournaras2014measuring,pournaras2014decentralized}. This research contributes a new application of I-EPOS to fog service placement and provides fundamental insights on how the provisioning of IoT services can be modeled as a multiple-choice combinatorial optimization problem.
\par As a result of the plan generation step, each agent comes with a certain number of possible plans and their corresponding cost. In the second step, all agents collaborate to choose their selected plans from these possible plans in terms of two objectives; \small{MIN-COST} \normalsize and \small{MIN-VAR}\normalsize. Agents are self-organized in a tree overlay topology as a way to structure their interactions with which they perform a cooperative optimization. The optimization is performed by a set of consecutive learning iterations consisting of two phases, the bottom-up (leaves to root) and top-down (root to leaves). At each iteration, agents change their selected plans combining the two objectives in a weighted sum of costs as equation (\ref{eq:18}) to reduce the costs compared to the previous iteration. The weighted summation can be used to make several trade-offs and provide multiple levels of QoS.
\begin{equation}
\lambda L^{t}+(1-\lambda) G^{t}
\label{eq:18}
\end{equation}
where $\lambda$ $\epsilon$ $[0,1]$. The higher the value of the weight, the stronger the preference towards minimizing the corresponding objective. When $\lambda = 1$ agents ignore the global cost while they maximize the local cost minimization. The cost functions take as an argument the global plan at the iteration $t-1$, which is the sum of all utilization plans of the agents in the network. The global cost function (\small{MIN-VAR} \normalsize objective) and the local cost function (\small{MIN-COST} \normalsize objective) with input all selected plans are formulated as follows:
\begin{align}
G^{t}& = \sigma(g^{t}),\; L^{t} = \min\frac{1}{N}\sum_{j=1}^{N}l(\delta_{j}^{(t)})\;\;, where\; G^{t},L^{t}\epsilon\,\mathbb{R} \end{align}
where $l(.)$ extracts the cost of the selected plan $\delta$ of the agent j at iteration t.
\par Regarding I-EPOS termination criteria, the system runtime completes when the global cost does not any longer change, or a certain number of iterations are performed. After termination, agents propagate their received requests to the selected hosts based on the selected plans. Accordingly, the hosts execute the requests while receiving new ones and perform the placement process again.

In terms of performance, earlier work demonstrates the computational and communication complexity of I-EPOS as well as its superior cost-effectiveness compared to state-of-the-art~\cite{pournaras2018decentralized}: (i) Low communication cost achieved via efficient information propagation in a network topology self-organized in a tree structure. (ii) Monotonic rapid learning performance in very few learning iterations. In terms of optimality, I-EPOS reaches solutions close to top- 3\% and above in optimization landscapes with over 1M of possible solutions. Recent findings expand the analysis with optimality profiles in large-scale networks~\cite{8875188}.

\section{Evaluation}\label{sec:evaluation}
\par Studying the proposed solutions in the context of IoT comes with several significant challenges~\cite{dastjerdi2016fog,svorobej2019simulating}. The scale and the complexity of this system make it infeasible to use a realistic IoT prototype~\cite{svorobej2019simulating,ficco2017pseudo}, while constructing a testbed, is complex, costly, and time-intensive. In such a context, mathematical modeling employs graphs to model the relationships between data centers~\cite{filiposka2014complex}, fog infrastructure~\cite{lera2018availability}, and load-balancing environments~\cite{zhang2010load}. Hence, in this research, various network topologies are modeled through three well-known graph models that consist of Barabasi-Albert (BA)~\cite{barabasi1999emergence}, Watts–Strogatz (WS)~\cite{watts1998collective}, and Erdos-Renyi (ER)~\cite{erdos1959random,van2016random}. 
\par
\begin{figure}[!htbp]
\centering
\includegraphics[clip, trim=2.0cm 7.0cm 2.1cm 12.5cm, width=\columnwidth]{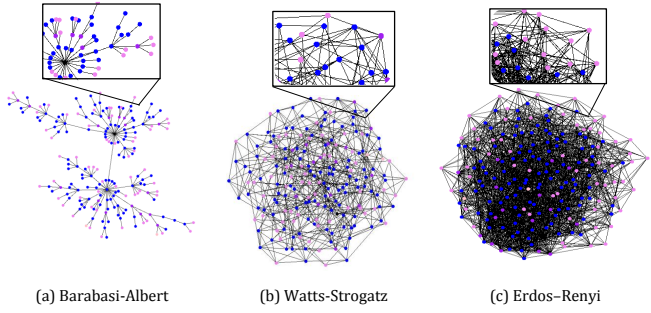}
\caption{Three graph types used to model network topology. These graphs illustrate a 200-node network. The color of the nodes is based on the processing capacity and their degree in the graph.}
\label{fig:graph}
\end{figure}

 Barabasi-Albert is a model for scale-free\footnote{The scale-free phenomenon declares that the degrees in real-world networks show an enormous amount of variability~\cite{schintler2003scale}.} networks such as the World Wide Web (w3), characterized by a highly heterogeneous degree distribution and high modularity (groups of the nodes that are more densely connected together than to the rest of the network). Erdos-Renyi model, known as a random network, has low heterogeneity, short average paths, and low clustering~\cite{sohn2017small,sole2004information}. Watts–Strogatz is a model for small-world\footnote{The small-world phenomenon states that distances in real-world networks are quite small~\cite{kleinberg1999small}.} networks which are very close structurally to social networks.  Fig.~\ref{fig:graph} shows the network graphs of the three selected models for a 200-node network.
Experimental evaluation is performed using a Java software that emulates a network of edge-to-cloud nodes. Besides, graph modeling and analysis are performed using a Java library, i.e., GraphStream\footnote{Available at: http://graphstream-project.org (last accessed: Jan 2021)}~\cite{dutot2007graphstream}.
%[clip, trim=0.0cm 1.2cm 12.1cm 0.3cm,,height=22.8cm

\begin{figure}[!htbp]
\centering
\includegraphics[clip, trim=2.8cm 19.0cm 6.2cm 2.8cm, width=\columnwidth]{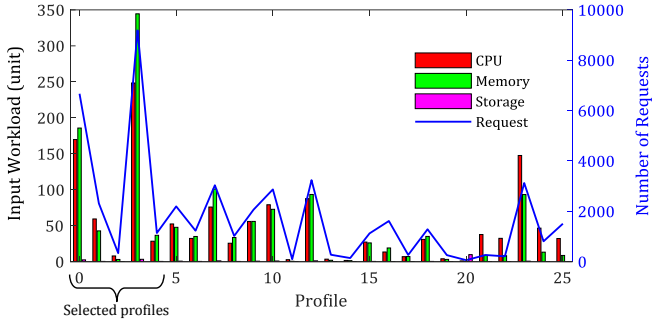}
\caption{26 profiles of Google workload~\cite{reiss2011google} in the form of 5-minute profiles. The first 5 profiles are selected as the input workload for the evaluation. Each profile is defined with four values that consist of: CPU load, memory load, storage load, and the number of requests.}
\label{fig:workload}
\end{figure}

\par As the input workload, the Google cluster trace\footnote{Available at: https://commondatastorage.googleapis.com/clusterdata-2011-2 (last accessed: Jan 2021).}~\cite{clusterdata:Wilkes2011} is used, which contains the data collected from a variety of input workloads on 12500 machines for 30 days. Fig.~\ref{fig:workload} displays the workload and corresponding requests during the first 130 minutes of the trace. Each 5-min period is referred to as a profile. 
In order to have a comprehensive evaluation, the profiles 0-4 have been tested, in which the input workload is highly variable. For the experiments, EPOS Fog execution time interval equals to profile duration ($\mu$ = 5 minutes). However, it is interesting to examine the performance of EPOS Fog with different values of these two parameters.

\par The number of possible plans per agent and number of iterations are set to 20 and 40, respectively. Each service request is accompanied by a set of resource requirements that consist of CPU, memory, and storage demands. Similar to the Google trace, the exact number of CPU cores and bytes of memory are unavailable; instead, resource demand information is provided in normalized units (measurements are expressed relative to the capacity of the most powerful machine).
The capacity values of the cloud and network nodes in the form of (CPU, memory, storage) are set to (400, 500, 200) and (704.0, 792.5, 313.5), which are specified in such a way that there is enough capacity available in the network to respond to all requests received.
Note that because the Google trace does not contain any value as a service deadline, 22 delay-sensitive services~\cite{byers2017architectural,wang2017virtual} are considered as a variety of IoT services, and their deadline values are associated with all input service requests, listed in Table \ref{tab:service-deadline}.

\begin{table}[!htbp]
\setlength\tabcolsep{0pt}
\scriptsize{
\centering
\caption {IoT Services and Corresponding Deadlines~\cite{byers2017architectural,wang2017virtual}}
\label{tab:service-deadline}
\smallskip
\begin{tabular*}{\columnwidth}{@{\extracolsep{\fill}}lr}
\hline
Service&Deadline\\
\hline
    Big data file download, off-line backup&100s\\
    YouTube, home automation, video surveillance&10s\\
    Web search, sensor readings&1s\\
    Interactive web site, smart building, analytics&100ms\\
    Broadcast&50ms\\
    Web game&30ms\\
    Virtual reality, smart transportation, finance, accelerated video&10ms\\
    Health care&5ms\\
    Augmented reality&2-10ms\\
    Haptics, robotics, real-time manufacturing, self-driving&1ms\\
\hline
\end{tabular*}
}
\end{table}

\par The conducted experiments analyze the relationships between the evaluated approaches and several configuration parameters. (i) \emph{Network size (N)}: To study the scalability of the proposed work, different numbers of nodes are considered for the network: 200, 400, 600, 800, 1000. (ii) \emph{IoT workload distribution}: The workload distribution parameter determines the distribution of IoT requests over the network. However, the availability of openly available datasets about the distribution of requests in real IoT scenarios is scarce~\cite{amadeo2019fog}. Therefore, considering literature~\cite{fan2018application,barros2018iot}, this paper explores the effect of two distributions that consist of a random (denoted as Rand in the experimental results)~\cite{ccinlar2011probability} and a Beta distribution~\cite{johnson1995chapter} as Beta (2.0, 5.0) on the performance results. (iii) \emph{Host proximity (H)}: This parameter investigates the impact of the distance between sources (i.e., service requesters) and corresponding destinations (i.e., hosts) on the evaluated metrics. Different distances in terms of hop count include 1-hop (direct neighbors), 3-hop, and $\infty$-hop (unlimited). Note that the host proximity constraint is applied in selecting host nodes in the plan generation step (line 6 in Algorithm~\ref{alg:plan-generation}). (iv) \emph{Agent preference ($\lambda$)}: This aspect examines the impact of different $\lambda$ values in the interval [0,1] on the  global and local cost reduction.

\subsection{Strategies and evaluation metrics}\label{sec:strategy}
\par Three approaches are considered for evaluation and comparisons. (i) \emph{Cloud}: This approach assumes that the fog infrastructure is not available, and all services are sent directly to the cloud. (ii) \emph{First Fit}~\cite{brent1989efficient,skarlat2017}: In this approach, each node traces the latency of the communication link between itself and other nodes and sorts the list of its direct neighbor nodes. Then, upon the receipt of each request, the list is checked, and if any node in the list meets the request requirements, it is sent to that node. Otherwise, the request is propagated to the cloud. (iii) \emph{EPOS Fog}: The proposed approach outlined in section~\ref{sec:cooperative-service}.
\par In order to show how the proposed service placement approach meets the objectives, the following metrics are evaluated.
(i) Utilization variance: It measures the workload balance among the nodes. To establish precise measurements for the load-balance, the three parameters of CPU, memory, and overall (CPU along with memory) load are considered. (ii) Average utilization of the fog infrastructure: This criterion shows to what extent fog nodes are utilized and is determined as a ratio of the workload placed on the network resources to the capacity of the resources, averaged for all nodes in the network.
(iii) Average deadline violations: This metric indicates the ratio of the number of services whose deadlines have been violated. (iv) Average service execution delay: The difference between service deadline and its response time, measured as $|\tau_{i}-e_{i,j}|$. (v) Utilization variance error: This metric measures how far the predicted utilization variance (the results obtained by I-EPOS) is from the realized one (the results of applying the I-EPOS plans on the network), as $|\sigma_{\delta}-{\sigma}'_{\delta}|$.  This error originates from the unhosted services that cannot be served due to insufficient resources available in the node allocated.  Intuitively, the higher the imbalance in the network, the higher the error. Consequently, by tuning EPOS Fog to improve load-balancing, the utilization variance error can also decrease. To assess this hypothesis, this paper focuses on the relation between this error and the $\lambda$ parameter that regulates the trade-off between the local and global objectives based on which the plan selections are performed. Higher $\lambda$ values decrease the degree of freedom to choose the plans with lower variance, while distributing services mostly across local regions deployed close to data sources. As a result of increasing the number of high-load nodes, the likelihood of capacity constraints violation due to future deployments increases, thereby limiting the load-balancing potential.
\subsection{Results and discussion}\label{sec:result}
\par This section assesses the execution of service placement plans provided by EPOS Fog, First Fit, and Cloud approaches. %Due to space limitation, only the results for 200- and 400-node networks, as well as the first and second profiles, are shown.
%Lambda=0.0, diff CPU

\subsubsection{Utilization variance}\label{sec:res-uv}
\par
\begin{figure}[!htbp]
\centering
\includegraphics[clip, trim=3.6cm 19.2cm 4.4cm 3.1cm, width=\columnwidth]{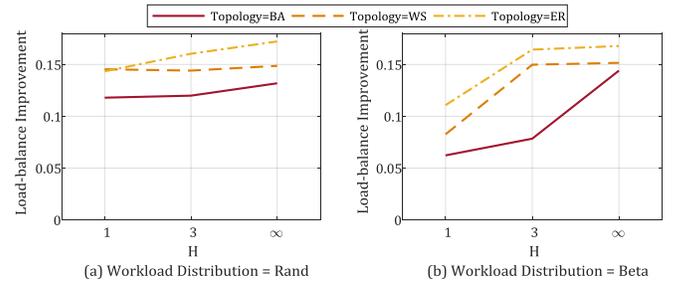}
\caption{Difference between overall (CPU along with memory) utilization variance of EPOS Fog and First Fit under varied parameters (Profile=1, N=400).}%Variance reduction
\label{fig:Ruv}
\end{figure}

 This studied aspect examines how well balanced the workload is distributed on the network. The Cloud approach does not perform any load-balancing by design, and therefore, it is excluded from this evaluation. Fig.~\ref{fig:Ruv} illustrates the difference between the utilization variance (i.e., reduction in utilization variance) of EPOS Fog and First Fit for a 400-node network. Figs.~\ref{fig:uv1}, \ref{fig:uv2}, and \ref{fig:uv3} in Appendix~\ref{Asec:plots} present the detailed figures on utilization variance.
\par In all scenarios, the utilization variance, i.e. global cost, in First Fit is between 40\% to 90\% higher than EPOS Fog. This is because, for First Fit, services are located on direct neighboring nodes where possible. Otherwise, they are forwarded to the cloud. In contrast, for EPOS Fog, the range of hosts is controlled by a host proximity constraint that can distribute services to a broader range of nodes. 
The difference between the utilization variance of the two approaches is higher by increasing the host proximity criterion from one to infinity. Larger proximity value allows hosting services on a wider number of nodes, which reduces the utilization variance in EPOS Fog.
\par With respect to EPOS Fog, the following observations can be made. In the case of different topologies, the utilization variance of WS is lower than BA up to 37\%, and the utilization variance of ER is lower than the other two topologies up to 45\%. This is due to the different characteristics of these topologies; short path lengths and low clustering measures support the higher balanced distribution of workload.
\par In general, for EPOS Fog, increasing the host proximity parameter from one to infinity decreases the utilization variance from 2\% to 40\%; the higher the degree of freedom to choose host nodes the more uniform the distribution of workload over the network. In around 90\% of the scenarios with random service distribution, the utilization variance is lower than the same scenarios with Beta distribution. This is because it is harder to achieve a balanced distribution when the input workload is not distributed uniformly.
Upon considering the workload distribution and host proximity together, it is observed that the difference of utilization variance between a random distribution scenario and the same scenario with Beta distribution increases with decreasing host proximity. In the case of 1-hop, the difference reaches 65\%. This is because, on the one hand, requests are not distributed uniformly with a Beta distribution, and on the other hand, as the host proximity value decreases, the range of the nodes that can be selected as a host becomes limited. However, this is not the case when there is no forwarding constraint (i.e., $\infty$-hop), which results in hosting the services on any distant nodes to achieve a higher balance. Note that these situations are only two cases among 18 configurations (i.e., 11\%).
\par It is worth noting that the utilization variance does not change significantly when comparing the results for networks with 200 and 400 nodes. This indicates that with increasing the number of nodes (with the constant workload and fixed network capacity), the workload balance remains at a similar level, indicating the scalability of the proposed approach given that the utilization variance in bounded to 0. 

% Cpu utilization + Lambda=0.0

\subsubsection{Utilization of the fog infrastructure}\label{sec:res-util}
\begin{figure}[!t]
\centering
\includegraphics[clip, trim=5.1cm 15.75cm 5.7cm 7.65cm, width=\columnwidth]{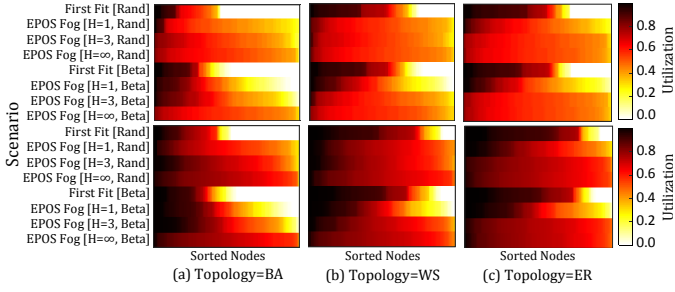}
\caption{Utilization of network resources for different scenarios under varied parameters (N=200, up: Profile=0, down: Profile=1).}
\label{fig:util}
\end{figure}
\par Fig.~\ref{fig:util} shows the utilization of network nodes for several scenarios. In each scenario, the nodes are sorted in descending order according to their utilization value.
For the Cloud approach, 100\% of placements are in the cloud node, and the fog resources are not utilized. Concerning First Fit, some nodes are used extensively (utilization is more than 90\%), while other nodes have very low load (utilization is less than 10\%). This is an artifact of the service placement strategy in First Fit: despite the free capacity in non-neighboring fog nodes, these available resources are not optimally utilized. EPOS Fog service placement employs fog resources more effectively, leading to reduced cloud utilization. For instance, regarding EPOS Fog [H=3, Rand], EPOS Fog [H=$\infty$, Rand], EPOS Fog [H=3, Beta], and EPOS Fog [H=$\infty$, Beta] in topology ER, almost all fog nodes have the utilization in the range [30\%, 80\%], while the utilization of cloud node is less than 10\%.
\par Given the increasing host proximity parameter in EPOS Fog, network nodes are allowed to select a broader range of fog nodes as host, and therefore, the utilization of these nodes increases while load-balances the network. For example, in the case of topology BA and profile 0, the utilization of nodes for EPOS Fog [H=1, Beta] ranges in [0\%, 85\%] while reaches to the range [20\%, 80\%] for EPOS Fog [H=3, Beta], and to [40\%, 70\%] for EPOS Fog [H=$\infty$, Beta]. This higher balanced distribution confirms the results of the previous subsection, i.e., variance reduction due to higher flexibility in host choices.
With respect to input profiles, although in both approaches nodes' utilization increases with a growing workload resulting from subsequent profiles. However, in contrast to First Fit in which nodes' utilization varies in the range [0\%, 100\%] for both first and second profiles, EPOS Fog distributes the workload more uniformly, which indicates a significant potential of EPOS Fog as a load-balancer under various input profiles. For instance, in EPOS Fog [H=$\infty$, Beta], nodes' utilization grows from the range [40\%, 65\%] in the first profile to the range [60\%, 80\%] in the second profile.
It is worth to be noticed that due to its random nature, ER topology provides a more uniform distribution of workload compared to other topologies, confirming the results of Section~\ref{sec:res-uv}, i.e., the topology strongly influences workload balance.
%Lambda=0.0

\subsubsection{Average deadline violations and service execution delay}\label{sec:res-sddl}
\par Because of the theoretically infinite resources in the cloud, requested services are executed immediately after submission and do not violate deadlines. Therefore, the Cloud approach is excluded from this evaluation. For the first profile, the average of deadline violations in First Fit is approximately 0.6, which is 1\% to 3\% higher than EPOS Fog. Moreover, this higher rate increases for the subsequent profiles.
Although different topologies have no considerable effect on this criterion in First Fit, in EPOS Fog the deadline violation for ER is slightly lower than WS, and for WS is lower than BA.
\par In order to study the response time of services in more detail, the average execution delay that services experience is assessed.
While the delay for the EPOS Fog and First Fit approaches in the first profile is approximately the same, in the second profile, this criterion is 1\% to 25\% higher in First Fit than in EPOS Fog. This is due to the fact that in First Fit, with increasing the number of requested services and decreased capacity in neighboring nodes, the forwarding of services to the cloud node increases, resulting in higher delay. 
In addition, the difference between the service delay in the two approaches changes across the different network topologies. For example, in profile 1, the service delay difference in the ER network is up to 25\% while in the BA network is up to 7\%.
Increasing the host proximity parameter results in 2\% to 17\% lower service delay in EPOS Fog compared to First Fit.
This is because, with a higher load-balance, the number of overloaded nodes decreases, thereby reducing the service delay and the probability of deadline violated. Moreover, it is interesting to know that even in the scenarios with a value of one for the proximity parameter (i.e., H=1), EPOS Fog provides from 1\% to 25\% lower delay than First Fit. That is because of the load-balancing strategy of EPOS Fog, which results in a reduced service delay. The reduction in service delay from EPOS Fog to First Fit is depicted in Fig.~\ref{fig:Rsd}. Detailed results are included in Appendix~\ref{Asec:plots} for more comprehensive comparisons.
\par Upon considering service execution delay and utilization results together, it is concluded that EPOS Fog provides both better fog utilization and lower service delay than First Fit. Note that the better performance even enhances in subsequent profiles. This is because, at the beginning, more resources are available, which makes the placement of requests easier for all strategies. However, in subsequent profiles, with the increasing number of requests, the placement has a higher impact on the utilization of the nodes.

\begin{figure}[!htbp]
\centering
\includegraphics[clip, trim=3.6cm 18.5cm 4.3cm 3.0cm, width=\columnwidth]{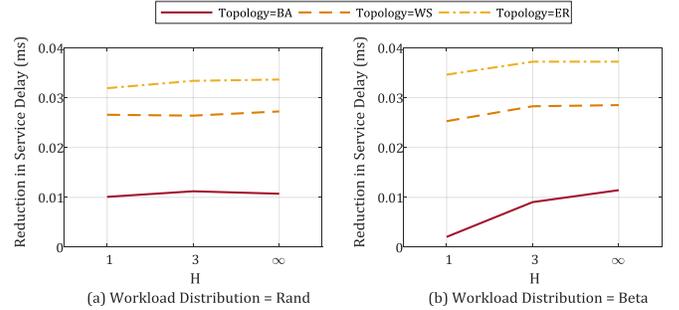}
\caption{Difference between service execution delay of EPOS Fog and First Fit under varied parameters (Profile=1, N=400).}
\label{fig:Rsd}
\end{figure}
%cpu utilization difference

\subsubsection{Utilization variance error}\label{sec:res-error}
\par
\begin{figure}[!htbp]
\centering
\includegraphics[trim=3.7cm 9.2cm 4.71cm 3.0cm, width=\columnwidth]{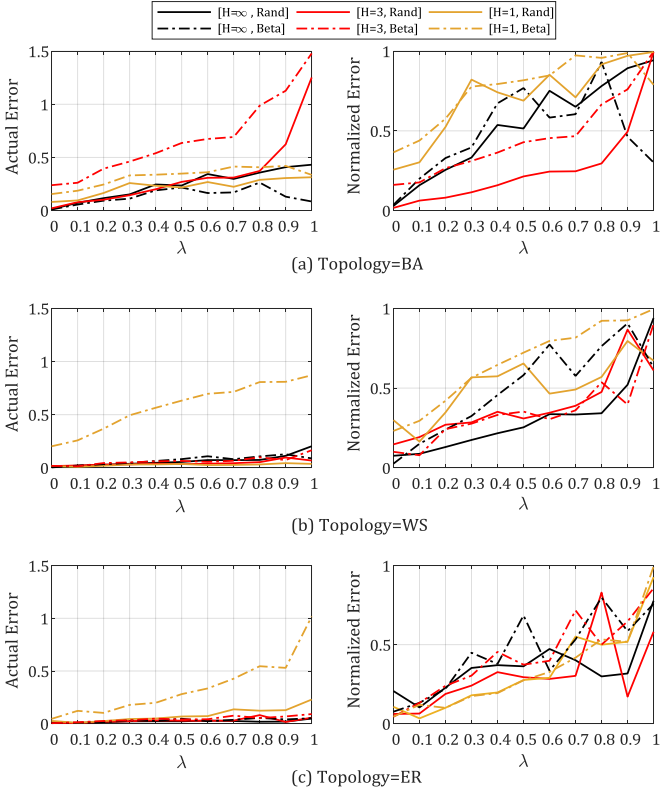}
\caption{Utilization variance error for different scenarios under varied $\lambda$ values (Profile=1, N=200).}
\label{fig:error}
\end{figure}

 Given the $\lambda$ values in the range [0, 1], Fig.~\ref{fig:error} evaluates the utilization variance error as the difference between the predicted utilization variance and the actual one. Note that the error is provided using the max-min normalized values and the actual values.
\par Generally, as $\lambda$ increases the error experienced increases. This is because the higher $\lambda$ values lead to agents preferring lower local cost plans, which results in a more overloaded network and a higher probability of high-load nodes. Consequently, the execution of I-EPOS plans in the unbalanced network increases the probability of capacity constraints violation in the overloaded nodes, and therefore prevents the realization of predicted variance. With respect to topology impact, the BA topology shows the highest error rate, and the ER topology presents the lowest error value for the same values of $\lambda$. This is because the ER topology, with short average paths and low clustering coefficients, provides higher load-balancing (as discussed in Section~\ref{sec:res-uv}) than BA, resulting in lower error. Fig.~\ref{fig:error} confirms the significant increase of the error rate for the scenarios with Beta service distribution and 1-host, 3-host proximity values that generally provide the lowest load-balance in comparison with other scenarios.
\par It is worth noting that by comparing the results obtained from the networks with 200 and 400 nodes, it is observed that by doubling N at a constant network capacity, the error rate is reduced up to 80\%. This is because the increasing number of nodes reduces the probability of high-load nodes to a high extent, and thus the predicted variance is significantly accurate.
\par In brief: when agents make plan choices in favor of their individual (local) objective (high $\lambda$ values), the collective (global) objective (i.e., utilization variance) is sacrificed and the network is more overloaded. As a result, the planned variance reduction deviates more from the actual one. Reward mechanisms are means to encourage agents to change the choices of $\lambda$ as well as their selected plan in line with a preferred objective. Employing various incentivization mechanisms with respect to the location of the users generating the service requests is the subject of future work.
\subsubsection{Summary of findings}\label{sec:summary}
\par A summary of the key findings in the performed experiments is given below.
(i) EPOS Fog outperforms other approaches in both (i) minimizing the cost of service execution (Fig.~\ref{fig:Rsd}) to improve the QoS and (ii) load-balancing of input workload (Figs.~\ref{fig:Ruv},~\ref{fig:uv1},~\ref{fig:uv2}, and~\ref{fig:uv3}) to enhance resource utilization and prevent peak load situations. (ii) EPOS Fog better utilizes edge-to-cloud nodes (Fig.~\ref{fig:util}) to allocate the resources effectively and reduce data traffic over the network. (iii) Even though the deadlines in EPOS Fog have slightly lower violated rates than First Fit, the delays in service execution are lower in the range 1\% to 25\% in EPOS Fog (Figs.~\ref{fig:Rsd},~\ref{fig:sd1}, and~\ref{fig:sd2}) compared to First Fit.
(iv) For EPOS Fog, an increasing number of agents (i.e., nodes) in a fixed network capacity decreases global cost and lowers utilization variance error as well as deadline violation, indicating the scalability of the proposed approach. The same results are valid for an increasing proximity of the workload redistribution in terms of number of hops. 
(v) Concerning EPOS Fog, workload distribution and network topology have the potential to improve the performance even further. Topologies with short paths and low clustering measures such as ER, and uniform workload distributions such as random result in better overall performance.
(vi) Planning the utilization of the network is more effective (lower utilization variance errors for lower $\lambda$ values) when prioritizing system-wide optimization over the optimization of local objectives (Fig.~\ref{fig:error}).
\par In summary, the advantages of EPOS Fog can be observed under various input workloads and experimental scenarios due to its flexibility and better exploration of the computation resources within the fog continuum.

\section{Conclusion and Future Work}\label{sec:conclusion}
\par Resource provisioning in the evolving IoT infrastructure is crucial for tackling the limitations in cloud-based technologies while meeting a broad range of IoT services requirements. This paper studies how the optimization of IoT service placement using \small{MIN-VAR} \normalsize and \small{MIN-COST} \normalsize objectives improves the performance of IoT services, such as response time, and obtains a balanced distribution of workload while utilizing resources on the network edges. The proposed approach, EPOS Fog, introduces a local plan generation mechanism, and employees I-EPOS, a cooperative plan selection methodology, for the IoT service placement. While the distributed load-balancing resource allocation increases system robustness, the objectives can be extended, e.g. energy-saving or monetary costs.
\par The proposed method demonstrates that a decentralized management of edge/fog computing is feasible and can be an enabler for a rich ecosystem of fog applications. The extensive experimental findings using real-world input profiles on various networks confirm that EPOS Fog, via a better utilization of edge-to-cloud nodes provides a higher QoS and more balanced distribution of workload over the network, compared to the First Fit and Cloud approaches. These results, under several experimental scenarios, confirm the scalability of EPOS Fog. 
\par Future work includes the mobility of load generators and an improved QoS using social information, such as users' profile, in such a context. Another aspect to study is delay-tolerant IoT services along with delay-sensitive ones.

%DECISIONS ARE LOCAL
% if have a single appendix:
%\appendix[Proof of the Zonklar Equations]
% or
%\appendix  % for no appendix heading
% do not use \section anymore after \appendix, only \section*
% is possibly needed

% use appendices with more than one appendix
% then use \section to start each appendix
% you must declare a \section before using any
% \subsection or using \label (\appendices by itself
% starts a section numbered zero.)
%

\appendices
\section{Evaluation Results in Detail}\label{Asec:plots}
\par
\begin{figure}[!htb]
\centering
\includegraphics[trim=3.5cm 10cm 8.4cm 3.3cm, width=\columnwidth]{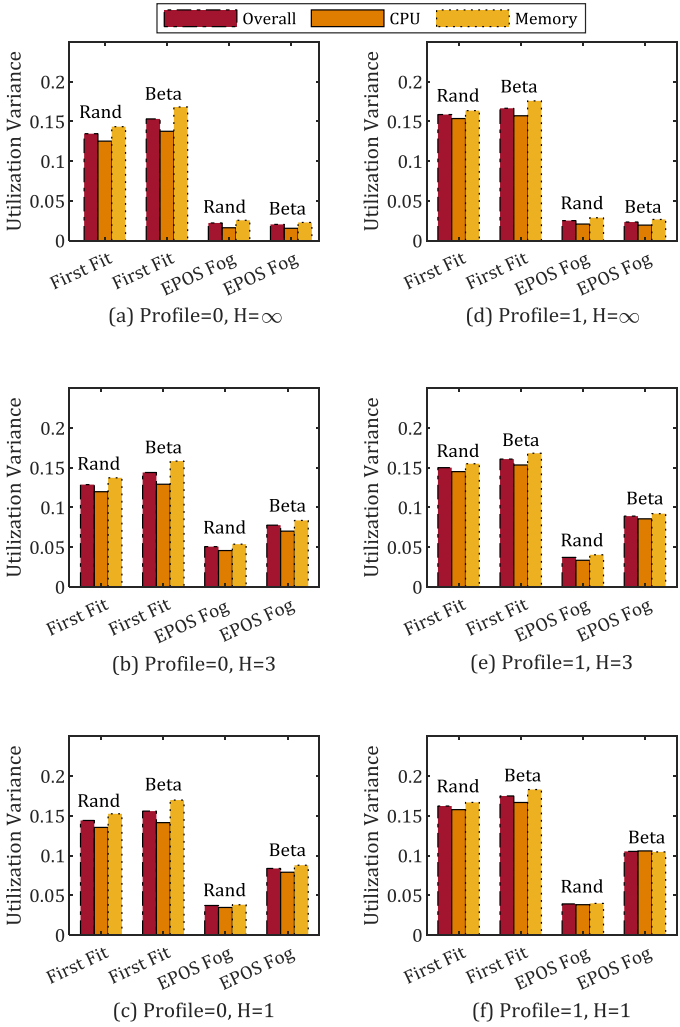}
\caption{Utilization variance for different scenarios under varied parameters (N=400, Topology=BA).}
\label{fig:uv1}
\end{figure}

\begin{figure}[!htbp]
\centering
\includegraphics[trim=3.5cm 10cm 8.4cm 3.2cm, width=\columnwidth]{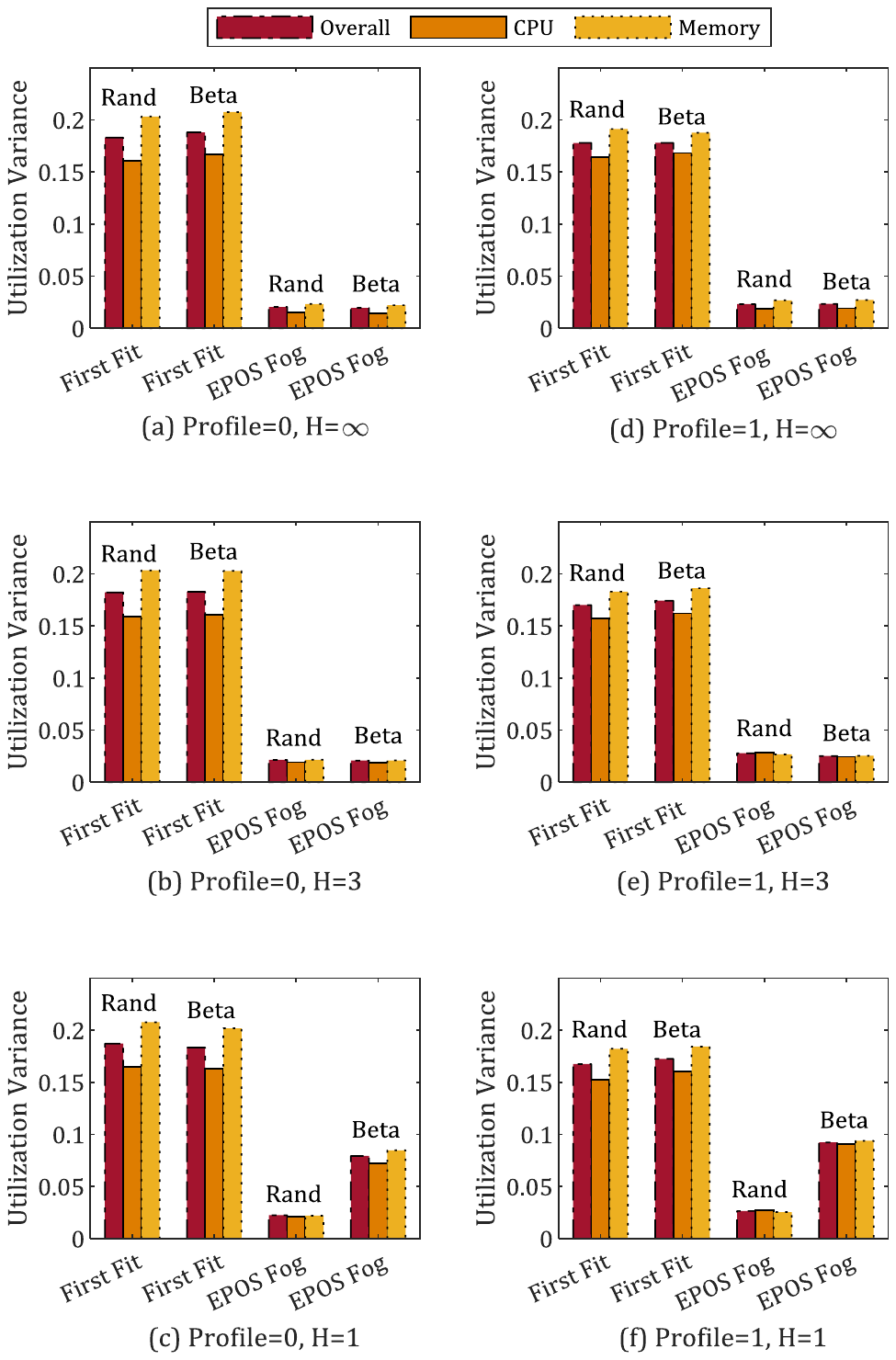}
\caption{Utilization variance for different scenarios under varied parameters (N=400, Topology=WS).}
\label{fig:uv2}
\end{figure}

\begin{figure}[!htbp]
\centering
\includegraphics[trim=3.5cm 10cm 8.4cm 3.2cm, width=\columnwidth]{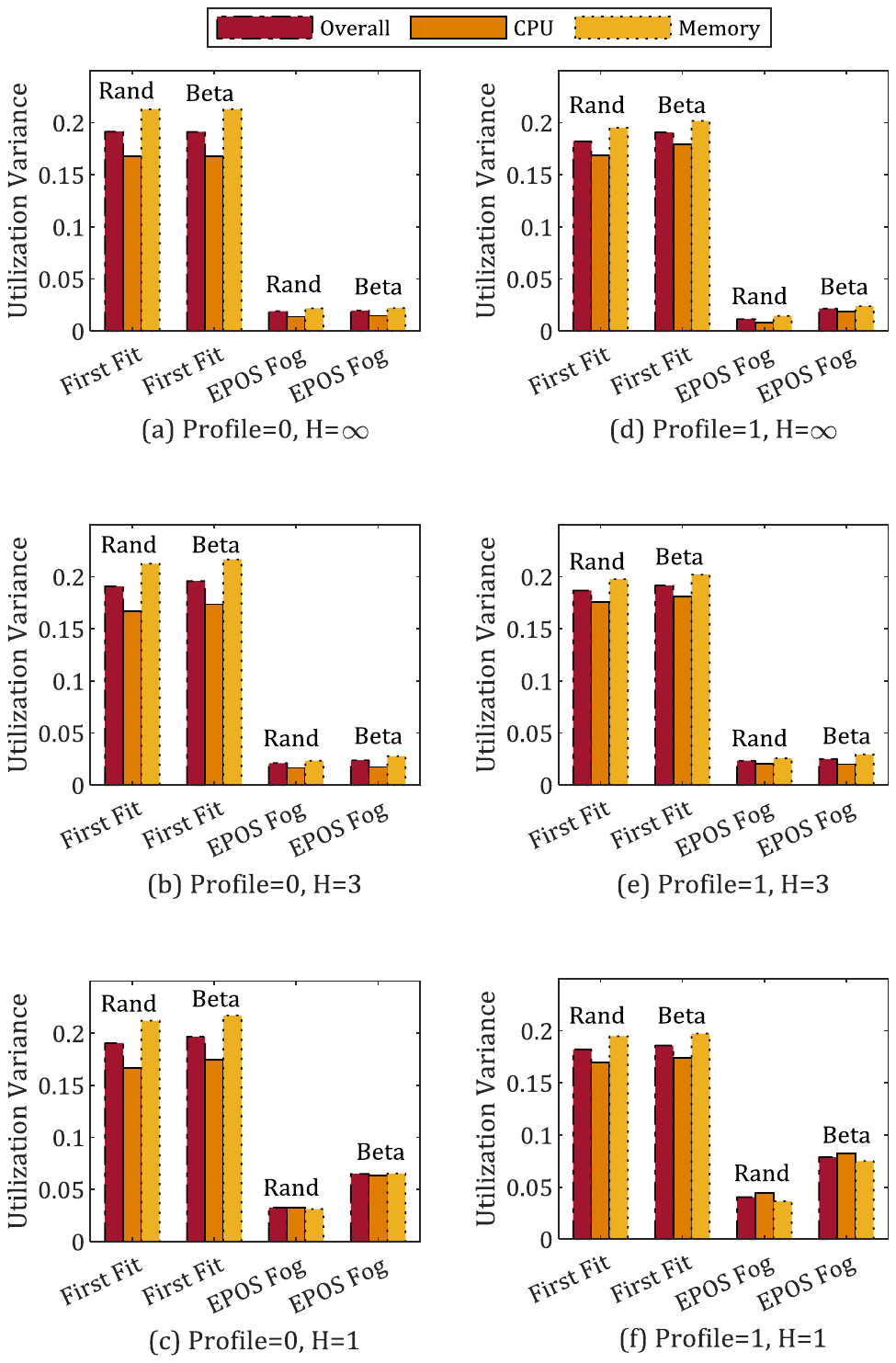}
\caption{Utilization variance for different scenarios under varied parameters (N=400, Topology=ER).}
\label{fig:uv3}
\end{figure}

\begin{figure}
\centering
\includegraphics[trim=3.5cm 10cm 8.4cm 3.2cm, width=\columnwidth]{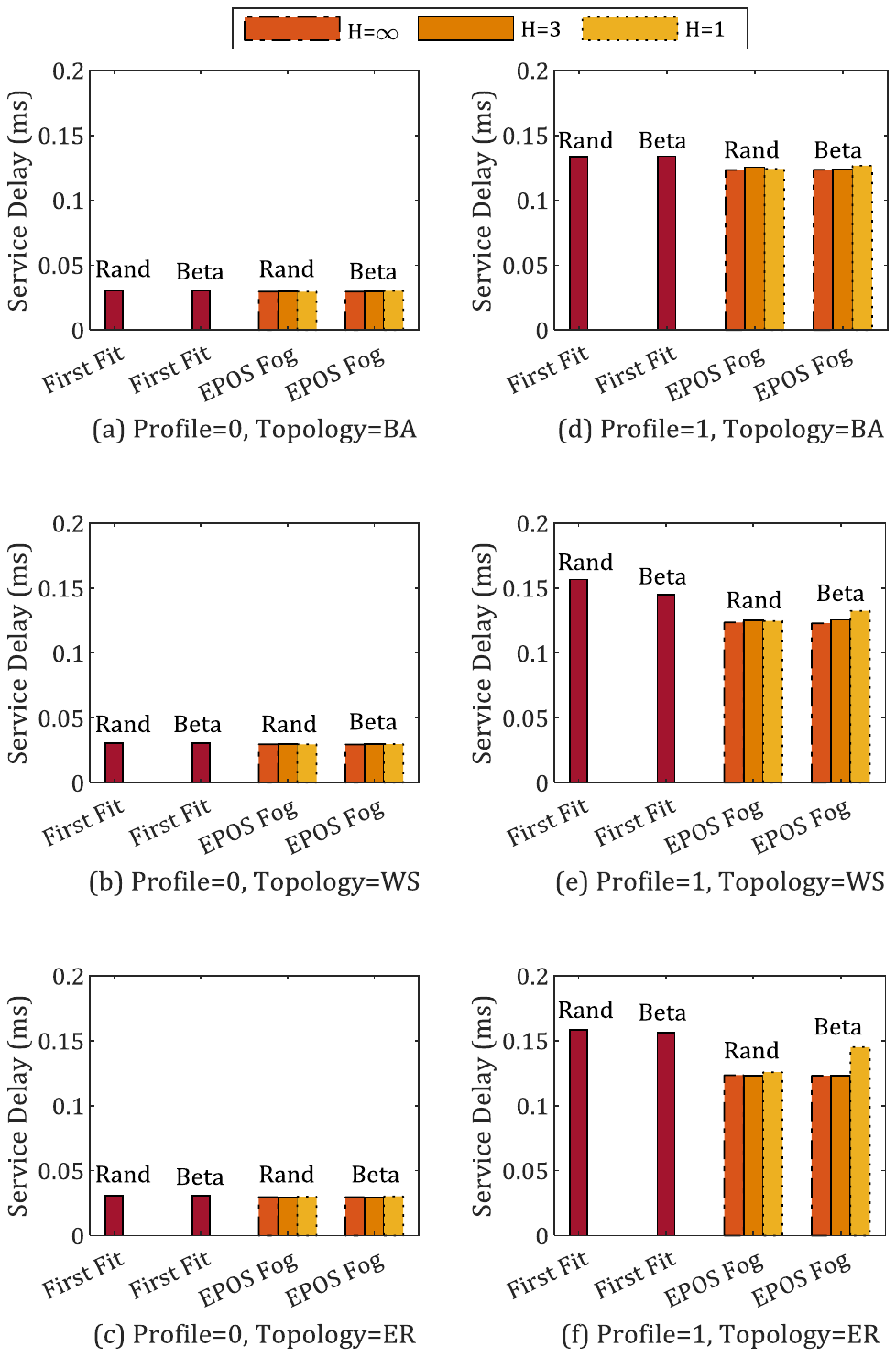}
\caption{Service execution delay for different scenarios under varied parameters (N=200).}
\label{fig:sd1}
\end{figure}

\begin{figure}
\centering
\includegraphics[trim=3.5cm 10cm 8.4cm 3.2cm, width=\columnwidth]{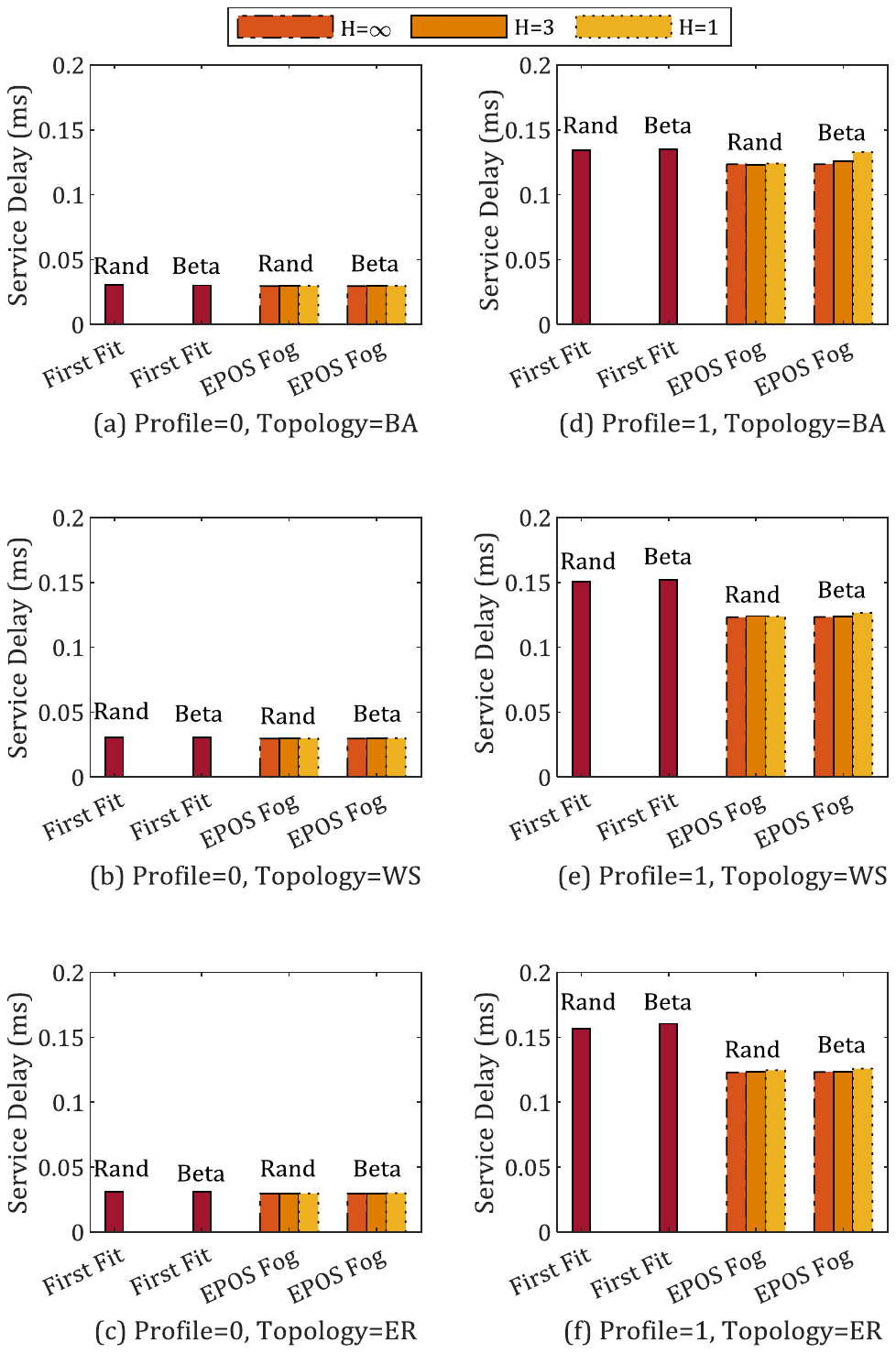}
\caption{Service execution delay for different scenarios under varied parameters (N=400).}
\label{fig:sd2}
\end{figure}

 Figs.~\ref{fig:uv1},~\ref{fig:uv2}, and~\ref{fig:uv3} illustrate the measurements of utilization variance and Figs.~\ref{fig:sd1} and~\ref{fig:sd2} show the service execution delay in detail. Results are illustrated considering the studied aspects: input profiles, host proximity constraint, network topology, and network size.
%\end{appendices}

%\section{Proof of the First Zonklar Equation}
% you can choose not to have a title for an appendix
% if you want by leaving the argument blank
%\section{}
%Appendix two text goes here.

% use section* for acknowledgement
\section*{Acknowledgment}
This research was financially supported by the government Ministry of Science, Research and Technology of the Islamic Republic of Iran. It was also partially supported by EPFL and ETH Zurich. The authors would like to thank Professor Dr. Dimitrios Kyritsis from ICT for Sustainable Manufacturing (ICT4SM) Group, EPFL, and Professor Dr. Dirk Helbing from Computational Social Science (CSS) Group, ETH Zurich, who provided simulation facilities and their support on this project greatly assisted the research. We would also like to thank Farzam Fanitabasi, PhD candidate at ETH Zurich, for his assistance with employing I-EPOS.

% Can use something like this to put references on a page
% by themselves when using endfloat and the captionsoff option.
\ifCLASSOPTIONcaptionsoff
  \newpage
\fi

\bibliographystyle{IEEEtran}
\bibliography{mybib}

% Generated by IEEEtran.bst, version: 1.14 (2015/08/26)
\begin{thebibliography}{10}
\providecommand{\url}[1]{#1}
\csname url@samestyle\endcsname
\providecommand{\newblock}{\relax}
\providecommand{\bibinfo}[2]{#2}
\providecommand{\BIBentrySTDinterwordspacing}{\spaceskip=0pt\relax}
\providecommand{\BIBentryALTinterwordstretchfactor}{4}
\providecommand{\BIBentryALTinterwordspacing}{\spaceskip=\fontdimen2\font plus
\BIBentryALTinterwordstretchfactor\fontdimen3\font minus
  \fontdimen4\font\relax}
\providecommand{\BIBforeignlanguage}[2]{{%
\expandafter\ifx\csname l@#1\endcsname\relax
\typeout{** WARNING: IEEEtran.bst: No hyphenation pattern has been}%
\typeout{** loaded for the language `#1'. Using the pattern for}%
\typeout{** the default language instead.}%
\else
\language=\csname l@#1\endcsname
\fi
#2}}
\providecommand{\BIBdecl}{\relax}
\BIBdecl

\bibitem{buyya2019fog}
\BIBentryALTinterwordspacing
R.~Buyya and S.~Srirama, \emph{Fog and Edge Computing: Principles and
  Paradigms}, ser. Wiley Series on Parallel and Distributed Computing.\hskip
  1em plus 0.5em minus 0.4em\relax Wiley, 2019. [Online]. Available:
  \url{https://books.google.com/books?id=cdSvtQEACAAJ}
\BIBentrySTDinterwordspacing

\bibitem{bonomi2014fog}
F.~Bonomi, R.~Milito, P.~Natarajan, and J.~Zhu, ``Fog computing: A platform for
  internet of things and analytics,'' in \emph{Big data and internet of things:
  A roadmap for smart environments}.\hskip 1em plus 0.5em minus 0.4em\relax
  Springer, 2014, pp. 169--186.

\bibitem{Nezami2019Internet}
Z.~Nezami and K.~Zamanifar, ``Internet of things/internet of everything:
  structure and ingredients,'' \emph{IEEE Potentials}, vol.~38, no.~2, pp.
  12--17, March 2019.

\bibitem{chiang2016fog}
M.~Chiang and T.~Zhang, ``Fog and iot: An overview of research opportunities,''
  \emph{IEEE Internet of Things Journal}, vol.~3, no.~6, pp. 854--864, 2016.

\bibitem{verma2016real}
M.~Verma, N.~Bhardwaj, and A.~K. Yadav, ``Real time efficient scheduling
  algorithm for load balancing in fog computing environment,'' \emph{Int. J.
  Inf. Technol. Comput. Sci}, vol.~8, no.~4, pp. 1--10, 2016.

\bibitem{verma2015architecture}
M.~Verma and N.~B. A.~K. Yadav, ``An architecture for load balancing techniques
  for fog computing environment,'' \emph{International Journal of Computer
  Science and Communication}, vol.~8, no.~2, pp. 43--49, 2015.

\bibitem{yousefpour2019all}
A.~Yousefpour, C.~Fung, T.~Nguyen, K.~Kadiyala, F.~Jalali, A.~Niakanlahiji,
  J.~Kong, and J.~P. Jue, ``All one needs to know about fog computing and
  related edge computing paradigms: A complete survey,'' \emph{Journal of
  Systems Architecture}, 2019.

\bibitem{openfog2017openfog}
O.~C. A.~W. Group \emph{et~al.}, ``Openfog reference architecture for fog
  computing,'' \emph{OPFRA001}, vol. 20817, p. 162, 2017.

\bibitem{brogi2017}
A.~Brogi and S.~Forti, ``Qos-aware deployment of iot applications through the
  fog,'' \emph{IEEE Internet of Things Journal}, vol.~4, no.~5, pp. 1185--1192,
  2017.

\bibitem{colistra2015task}
G.~Colistra, ``Task allocation in the internet of things,'' Ph.D. dissertation,
  Universita'degli Studi di Cagliari, 2015.

\bibitem{yousefpour2019fogplan}
A.~Yousefpour, A.~Patil, G.~Ishigaki, I.~Kim, X.~Wang, H.~C. Cankaya, Q.~Zhang,
  W.~Xie, and J.~P. Jue, ``Fogplan: A lightweight qos-aware dynamic fog service
  provisioning framework,'' \emph{IEEE Internet of Things Journal}, 2019.

\bibitem{kumar2014distributed}
N.~Kumar, S.~Agarwal, T.~Zaidi, and V.~Saxena, ``A distributed load-balancing
  scheme based on a complex network model of cloud servers,'' \emph{ACM SIGSOFT
  Software Engineering Notes}, vol.~39, no.~6, pp. 1--6, 2014.

\bibitem{song2017approach}
Y.~Song, S.~S. Yau, R.~Yu, X.~Zhang, and G.~Xue, ``An approach to qos-based
  task distribution in edge computing networks for iot applications,'' in
  \emph{2017 IEEE International Conference on Edge Computing (EDGE)}.\hskip 1em
  plus 0.5em minus 0.4em\relax IEEE, 2017, pp. 32--39.

\bibitem{khattak2019utilization}
H.~A. Khattak, H.~Arshad, S.~ul~Islam, G.~Ahmed, S.~Jabbar, A.~M. Sharif, and
  S.~Khalid, ``Utilization and load balancing in fog servers for health
  applications,'' \emph{EURASIP Journal on Wireless Communications and
  Networking}, vol. 2019, no.~1, p.~91, 2019.

\bibitem{rahmani2015smart}
A.-M. Rahmani, N.~K. Thanigaivelan, T.~N. Gia, J.~Granados, B.~Negash,
  P.~Liljeberg, and H.~Tenhunen, ``Smart e-health gateway: Bringing
  intelligence to internet-of-things based ubiquitous healthcare systems,'' in
  \emph{2015 12th Annual IEEE Consumer Communications and Networking Conference
  (CCNC)}.\hskip 1em plus 0.5em minus 0.4em\relax IEEE, 2015, pp. 826--834.

\bibitem{gia2019exploiting}
T.~N. Gia and M.~Jiang, ``Exploiting fog computing in health monitoring,''
  \emph{Fog and Edge Computing: Principles and Paradigms}, pp. 291--318, 2019.

\bibitem{banerjee2017multi}
S.~Banerjee and J.~P. Hecker, ``A multi-agent system approach to load-balancing
  and resource allocation for distributed computing,'' in \emph{First Complex
  Systems Digital Campus World E-Conference 2015}.\hskip 1em plus 0.5em minus
  0.4em\relax Springer, 2017, pp. 41--54.

\bibitem{8595391}
M.~{D'Angelo}, ``Decentralized self-adaptive computing at the edge,'' in
  \emph{2018 IEEE/ACM 13th International Symposium on Software Engineering for
  Adaptive and Self-Managing Systems (SEAMS)}, May 2018, pp. 144--148.

\bibitem{8875188}
J.~{Nikolic} and E.~{Pournaras}, ``Structural self-adaptation for decentralized
  pervasive intelligence,'' in \emph{2019 22nd Euromicro Conference on Digital
  System Design (DSD)}, Aug 2019, pp. 562--571.

\bibitem{santos2019resource}
J.~Santos, T.~Wauters, B.~Volckaert, and F.~De~Turck, ``Resource provisioning
  in fog computing: From theory to practice,'' \emph{Sensors}, vol.~19, no.~10,
  p. 2238, 2019.

\bibitem{alsaffar2016architecture}
A.~A. Alsaffar, H.~P. Pham, C.-S. Hong, E.-N. Huh, and M.~Aazam, ``An
  architecture of iot service delegation and resource allocation based on
  collaboration between fog and cloud computing,'' \emph{Mobile Information
  Systems}, vol. 2016, 2016.

\bibitem{fan2018application}
Q.~Fan and N.~Ansari, ``Application aware workload allocation for edge
  computing-based iot,'' \emph{IEEE Internet of Things Journal}, vol.~5, no.~3,
  pp. 2146--2153, 2018.

\bibitem{pournaras2018decentralized}
E.~Pournaras, P.~Pilgerstorfer, and T.~Asikis, ``Decentralized collective
  learning for self-managed sharing economies,'' \emph{ACM Transactions on
  Autonomous and Adaptive Systems (TAAS)}, vol.~13, no.~2, p.~10, 2018.

\bibitem{cardellini2016optimal}
V.~Cardellini, V.~Grassi, F.~Lo~Presti, and M.~Nardelli, ``Optimal operator
  placement for distributed stream processing applications,'' in
  \emph{Proceedings of the 10th ACM International Conference on Distributed and
  Event-based Systems}.\hskip 1em plus 0.5em minus 0.4em\relax ACM, 2016, pp.
  69--80.

\bibitem{zhan2015cloud}
Z.-H. Zhan, X.-F. Liu, Y.-J. Gong, J.~Zhang, H.~S.-H. Chung, and Y.~Li, ``Cloud
  computing resource scheduling and a survey of its evolutionary approaches,''
  \emph{ACM Computing Surveys (CSUR)}, vol.~47, no.~4, p.~63, 2015.

\bibitem{leitner2012cost}
P.~Leitner, W.~Hummer, B.~Satzger, C.~Inzinger, and S.~Dustdar,
  ``Cost-efficient and application sla-aware client side request scheduling in
  an infrastructure-as-a-service cloud,'' in \emph{2012 IEEE Fifth
  International Conference on Cloud Computing}.\hskip 1em plus 0.5em minus
  0.4em\relax IEEE, 2012, pp. 213--220.

\bibitem{skarlat2017}
O.~Skarlat, M.~Nardelli, S.~Schulte, M.~Borkowski, and P.~Leitner, ``Optimized
  iot service placement in the fog,'' \emph{Service Oriented Computing and
  Applications}, vol.~11, no.~4, pp. 427--443, 2017.

\bibitem{souza2016handling}
V.~B. C.~d. Souza, W.~Ram{\'\i}rez, X.~Masip-Bruin, E.~Mar{\'\i}n-Tordera,
  G.~Ren, and G.~Tashakor, ``Handling service allocation in combined fog-cloud
  scenarios,'' in \emph{Communications (ICC), 2016 IEEE International
  Conference on}.\hskip 1em plus 0.5em minus 0.4em\relax IEEE, 2016, pp. 1--5.

\bibitem{fadahunsi2019locality}
O.~Fadahunsi and M.~Maheswaran, ``Locality sensitive request distribution for
  fog and cloud servers,'' \emph{Service Oriented Computing and Applications},
  pp. 1--14, 2019.

\bibitem{xia2018combining}
Y.~Xia, X.~Etchevers, L.~Letondeur, T.~Coupaye, and F.~Desprez, ``Combining
  hardware nodes and software components ordering-based heuristics for
  optimizing the placement of distributed iot applications in the fog,'' in
  \emph{Proceedings of the 33rd Annual ACM Symposium on Applied
  Computing}.\hskip 1em plus 0.5em minus 0.4em\relax ACM, 2018, pp. 751--760.

\bibitem{skarlat2017towards}
O.~Skarlat, M.~Nardelli, S.~Schulte, and S.~Dustdar, ``Towards qos-aware fog
  service placement,'' in \emph{Fog and Edge Computing (ICFEC), 2017 IEEE 1st
  International Conference on}.\hskip 1em plus 0.5em minus 0.4em\relax IEEE,
  2017, pp. 89--96.

\bibitem{tran2019task}
M.-Q. Tran, D.~T. Nguyen, V.~A. Le, D.~H. Nguyen, and T.~V. Pham, ``Task
  placement on fog computing made efficient for iot application provision,''
  \emph{Wireless Communications and Mobile Computing}, vol. 2019, 2019.

\bibitem{deng2016optimal}
R.~Deng, R.~Lu, C.~Lai, T.~H. Luan, and H.~Liang, ``Optimal workload allocation
  in fog-cloud computing toward balanced delay and power consumption,''
  \emph{IEEE Internet of Things Journal}, vol.~3, no.~6, pp. 1171--1181, 2016.

\bibitem{naha2020deadline}
R.~K. Naha, S.~Garg, A.~Chan, and S.~K. Battula, ``Deadline-based dynamic
  resource allocation and provisioning algorithms in fog-cloud environment,''
  \emph{Future Generation Computer Systems}, vol. 104, pp. 131--141, 2020.

\bibitem{naha2021multi}
R.~K. Naha and S.~Garg, ``Multi-criteria--based dynamic user behaviour--aware
  resource allocation in fog computing,'' \emph{ACM Transactions on Internet of
  Things}, vol.~2, no.~1, pp. 1--31, 2021.

\bibitem{kapsalis2017cooperative}
A.~Kapsalis, P.~Kasnesis, I.~S. Venieris, D.~I. Kaklamani, and C.~Z.
  Patrikakis, ``A cooperative fog approach for effective workload balancing,''
  \emph{IEEE Cloud Computing}, vol.~4, no.~2, pp. 36--45, 2017.

\bibitem{xu2018dynamic}
X.~Xu, S.~Fu, Q.~Cai, W.~Tian, W.~Liu, W.~Dou, X.~Sun, and A.~X. Liu, ``Dynamic
  resource allocation for load balancing in fog environment,'' \emph{Wireless
  Communications and Mobile Computing}, vol. 2018, 2018.

\bibitem{8762052}
B.~{Donassolo}, I.~{Fajjari}, A.~{Legrand}, and P.~{Mertikopoulos}, ``Load
  aware provisioning of iot services on fog computing platform,'' in \emph{ICC
  2019 - 2019 IEEE International Conference on Communications (ICC)}, May 2019,
  pp. 1--7.

\bibitem{feo1995greedy}
T.~A. Feo and M.~G. Resende, ``Greedy randomized adaptive search procedures,''
  \emph{Journal of global optimization}, vol.~6, no.~2, pp. 109--133, 1995.

\bibitem{8931659}
J.~{Zhang}, H.~{Guo}, J.~{Liu}, and Y.~{Zhang}, ``Task offloading in vehicular
  edge computing networks: A load-balancing solution,'' \emph{IEEE Transactions
  on Vehicular Technology}, vol.~69, no.~2, pp. 2092--2104, 2020.

\bibitem{babou2020hierarchical}
C.~S.~M. Babou, D.~Fall, S.~Kashihara, Y.~Taenaka, M.~H. Bhuyan, I.~Niang, and
  Y.~Kadobayashi, ``Hierarchical load balancing and clustering technique for
  home edge computing,'' \emph{IEEE Access}, vol.~8, pp. 127\,593--127\,607,
  2020.

\bibitem{mouradian2017comprehensive}
C.~Mouradian, D.~Naboulsi, S.~Yangui, R.~H. Glitho, M.~J. Morrow, and P.~A.
  Polakos, ``A comprehensive survey on fog computing: State-of-the-art and
  research challenges,'' \emph{IEEE Communications Surveys \& Tutorials},
  vol.~20, no.~1, pp. 416--464, 2017.

\bibitem{bulkan2018load}
U.~Bulkan, T.~Dagiuklas, M.~Iqbal, K.~M.~S. Huq, A.~Al-Dulaimi, and
  J.~Rodriguez, ``On the load balancing of edge computing resources for on-line
  video delivery,'' \emph{IEEE Access}, vol.~6, pp. 73\,916--73\,927, 2018.

\bibitem{rajan2013survey}
R.~G. Rajan and V.~Jeyakrishnan, ``A survey on load balancing in cloud
  computing environments,'' \emph{International Journal of Advanced Research in
  Computer and Communication Engineering}, vol.~2, no.~12, pp. 4726--4728,
  2013.

\bibitem{deng2010heat}
Y.~Deng and R.~W. Lau, ``Heat diffusion based dynamic load balancing for
  distributed virtual environments,'' in \emph{Proceedings of the 17th ACM
  Symposium on Virtual Reality Software and Technology}.\hskip 1em plus 0.5em
  minus 0.4em\relax ACM, 2010, pp. 203--210.

\bibitem{gupta2017ifogsim}
H.~Gupta, A.~Vahid~Dastjerdi, S.~K. Ghosh, and R.~Buyya, ``ifogsim: A toolkit
  for modeling and simulation of resource management techniques in the internet
  of things, edge and fog computing environments,'' \emph{Software: Practice
  and Experience}, vol.~47, no.~9, pp. 1275--1296, 2017.

\bibitem{iorga2018fog}
M.~Iorga, L.~Feldman, R.~Barton, M.~J. Martin, N.~S. Goren, and C.~Mahmoudi,
  ``Fog computing conceptual model,'' Tech. Rep., 2018.

\bibitem{kaur2017container}
K.~Kaur, T.~Dhand, N.~Kumar, and S.~Zeadally, ``Container-as-a-service at the
  edge: Trade-off between energy efficiency and service availability at fog
  nano data centers,'' \emph{IEEE wireless communications}, vol.~24, no.~3, pp.
  48--56, 2017.

\bibitem{xiao2017qoe}
Y.~Xiao and M.~Krunz, ``Qoe and power efficiency tradeoff for fog computing
  networks with fog node cooperation,'' in \emph{INFOCOM 2017-IEEE Conference
  on Computer Communications, IEEE}.\hskip 1em plus 0.5em minus 0.4em\relax
  IEEE, 2017, pp. 1--9.

\bibitem{Serfozo1994}
\BIBentryALTinterwordspacing
R.~F. Serfozo, ``Little laws for utility processes and waiting times in
  queues,'' \emph{Queueing Systems}, vol.~17, no.~1, pp. 137--181, Mar 1994.
  [Online]. Available: \url{https://doi.org/10.1007/BF01158693}
\BIBentrySTDinterwordspacing

\bibitem{ningning2016fog}
S.~Ningning, G.~Chao, A.~Xingshuo, and Z.~Qiang, ``Fog computing dynamic load
  balancing mechanism based on graph repartitioning,'' \emph{China
  Communications}, vol.~13, no.~3, pp. 156--164, 2016.

\bibitem{liu2013data}
Z.~Liu, A.~Wierman, Y.~Chen, B.~Razon, and N.~Chen, ``Data center demand
  response: Avoiding the coincident peak via workload shifting and local
  generation,'' \emph{Performance Evaluation}, vol.~70, no.~10, pp. 770--791,
  2013.

\bibitem{zhang2017resilient}
H.~Zhang, J.~Zhang, W.~Bai, K.~Chen, and M.~Chowdhury, ``Resilient datacenter
  load balancing in the wild,'' in \emph{Proceedings of the Conference of the
  ACM Special Interest Group on Data Communication}.\hskip 1em plus 0.5em minus
  0.4em\relax ACM, 2017, pp. 253--266.

\bibitem{rockafellar2000optimization}
R.~T. Rockafellar, S.~Uryasev \emph{et~al.}, ``Optimization of conditional
  value-at-risk,'' \emph{Journal of risk}, vol.~2, pp. 21--42, 2000.

\bibitem{pournaras2019socio}
E.~Pournaras, S.~Jung, S.~Yadhunathan, H.~Zhang, and X.~Fang, ``Socio-technical
  smart grid optimization via decentralized charge control of electric
  vehicles,'' \emph{Applied Soft Computing}, vol.~82, p. 105573, 2019.

\bibitem{pournaras2017self}
E.~Pournaras, M.~Yao, and D.~Helbing, ``Self-regulating supply--demand
  systems,'' \emph{Future Generation Computer Systems}, vol.~76, pp. 73--91,
  2017.

\bibitem{pournaras2014measuring}
E.~Pournaras, M.~Vasirani, R.~E. Kooij, and K.~Aberer, ``Measuring and
  controlling unfairness in decentralized planning of energy demand,'' in
  \emph{2014 IEEE international energy conference (ENERGYCON)}.\hskip 1em plus
  0.5em minus 0.4em\relax IEEE, 2014, pp. 1255--1262.

\bibitem{pournaras2014decentralized}
------, ``Decentralized planning of energy demand for the management of
  robustness and discomfort,'' \emph{IEEE Transactions on Industrial
  Informatics}, vol.~10, no.~4, pp. 2280--2289, 2014.

\bibitem{dastjerdi2016fog}
A.~V. Dastjerdi and R.~Buyya, ``Fog computing: Helping the internet of things
  realize its potential,'' \emph{Computer}, vol.~49, no.~8, pp. 112--116, 2016.

\bibitem{svorobej2019simulating}
S.~Svorobej, P.~Takako~Endo, M.~Bendechache, C.~Filelis-Papadopoulos, K.~M.
  Giannoutakis, G.~A. Gravvanis, D.~Tzovaras, J.~Byrne, and T.~Lynn,
  ``Simulating fog and edge computing scenarios: An overview and research
  challenges,'' \emph{Future Internet}, vol.~11, no.~3, p.~55, 2019.

\bibitem{ficco2017pseudo}
M.~Ficco, C.~Esposito, Y.~Xiang, and F.~Palmieri, ``Pseudo-dynamic testing of
  realistic edge-fog cloud ecosystems,'' \emph{IEEE Communications Magazine},
  vol.~55, no.~11, pp. 98--104, 2017.

\bibitem{filiposka2014complex}
S.~Filiposka and C.~Juiz, ``Complex cloud datacenters,'' \emph{IERI Procedia},
  vol.~7, pp. 8--14, 2014.

\bibitem{lera2018availability}
I.~Lera, C.~Guerrero, and C.~Juiz, ``Availability-aware service placement
  policy in fog computing based on graph partitions,'' \emph{IEEE Internet of
  Things Journal}, vol.~6, no.~2, pp. 3641--3651, 2018.

\bibitem{zhang2010load}
Z.~Zhang and X.~Zhang, ``A load balancing mechanism based on ant colony and
  complex network theory in open cloud computing federation,'' in \emph{2010
  The 2nd International Conference on Industrial Mechatronics and Automation},
  vol.~2.\hskip 1em plus 0.5em minus 0.4em\relax IEEE, 2010, pp. 240--243.

\bibitem{barabasi1999emergence}
A.-L. Barab{\'a}si and R.~Albert, ``Emergence of scaling in random networks,''
  \emph{science}, vol. 286, no. 5439, pp. 509--512, 1999.

\bibitem{watts1998collective}
D.~J. Watts and S.~H. Strogatz, ``Collective dynamics of
  ‘small-world’networks,'' \emph{nature}, vol. 393, no. 6684, p. 440, 1998.

\bibitem{erdos1959random}
P.~Erd{\"o}s and A.~R{\'e}nyi, ``On random graphs, i,'' \emph{Publicationes
  Mathematicae (Debrecen)}, vol.~6, pp. 290--297, 1959.

\bibitem{van2016random}
R.~Van Der~Hofstad, \emph{Random graphs and complex networks}.\hskip 1em plus
  0.5em minus 0.4em\relax Cambridge university press, 2016, vol.~1.

\bibitem{schintler2003scale}
L.~A. Schintler, A.~Reggiani, R.~Kulkarni, and P.~Nijkamp, ``Scale-free
  phenomena in communication networks: A cross-atlantic comparison,'' 2003.

\bibitem{sohn2017small}
I.~Sohn, ``Small-world and scale-free network models for iot systems,''
  \emph{Mobile Information Systems}, vol. 2017, 2017.

\bibitem{sole2004information}
R.~V. Sol{\'e} and S.~Valverde, ``Information theory of complex networks: on
  evolution and architectural constraints,'' in \emph{Complex networks}.\hskip
  1em plus 0.5em minus 0.4em\relax Springer, 2004, pp. 189--207.

\bibitem{kleinberg1999small}
J.~Kleinberg, ``The small-world phenomenon: An algorithmic perspective,''
  Cornell University, Tech. Rep., 1999.

\bibitem{dutot2007graphstream}
A.~Dutot, F.~Guinand, D.~Olivier, and Y.~Pign{\'e}, ``Graphstream: A tool for
  bridging the gap between complex systems and dynamic graphs,'' in
  \emph{Emergent Properties in Natural and Artificial Complex Systems.
  Satellite Conference within the 4th European Conference on Complex Systems
  (ECCS'2007)}, 2007.

\bibitem{reiss2011google}
C.~Reiss, J.~Wilkes, and J.~L. Hellerstein, ``Google cluster-usage traces:
  format+ schema,'' \emph{Google Inc., White Paper}, pp. 1--14, 2011.

\bibitem{clusterdata:Wilkes2011}
J.~Wilkes, ``More {Google} cluster data,'' Google research blog, Nov. 2011,
  posted at
  \url{http://googleresearch.blogspot.com/2011/11/more-google-cluster-data.html}.

\bibitem{byers2017architectural}
C.~C. Byers, ``Architectural imperatives for fog computing: Use cases,
  requirements, and architectural techniques for fog-enabled iot networks,''
  \emph{IEEE Communications Magazine}, vol.~55, no.~8, pp. 14--20, 2017.

\bibitem{wang2017virtual}
W.~Wang, Y.~Zhao, M.~Tornatore, A.~Gupta, J.~Zhang, and B.~Mukherjee, ``Virtual
  machine placement and workload assignment for mobile edge computing,'' in
  \emph{Cloud Networking (CloudNet), 2017 IEEE 6th International Conference
  on}.\hskip 1em plus 0.5em minus 0.4em\relax IEEE, 2017, pp. 1--6.

\bibitem{amadeo2019fog}
M.~Amadeo, G.~Ruggeri, C.~Campolo, A.~Molinaro, V.~Loscr{\'\i}, and C.~T.
  Calafate, ``Fog computing in iot smart environments via named data
  networking: A study on service orchestration mechanisms,'' \emph{Future
  Internet}, vol.~11, no.~11, p. 222, 2019.

\bibitem{barros2018iot}
V.~A. Barros, J.~C. Estrella, L.~B. Prates, and S.~M. Bruschi, ``An iot-daas
  approach for the interoperability of heterogeneous sensor data sources,'' in
  \emph{Proceedings of the 21st ACM International Conference on Modeling,
  Analysis and Simulation of Wireless and Mobile Systems}.\hskip 1em plus 0.5em
  minus 0.4em\relax ACM, 2018, pp. 275--279.

\bibitem{ccinlar2011probability}
E.~{\c{C}}{\i}nlar, \emph{Probability and stochastics}.\hskip 1em plus 0.5em
  minus 0.4em\relax Springer Science \& Business Media, 2011, vol. 261.

\bibitem{johnson1995chapter}
N.~L. Johnson, S.~Kotz, and N.~Balakrishnan, ``Chapter 21: beta
  distributions,'' \emph{Continuous Univariate Distributions}, vol.~2, 1995.

\bibitem{brent1989efficient}
R.~P. Brent, ``Efficient implementation of the first-fit strategy for dynamic
  storage allocation,'' \emph{ACM Transactions on Programming Languages and
  Systems (TOPLAS)}, vol.~11, no.~3, pp. 388--403, 1989.

\end{thebibliography}

\end{document}